\documentclass[onecolumn, a4paper, 11p]{article}

\usepackage{booktabs}
\usepackage{tabularx}
\usepackage{paralist}
\usepackage[most]{tcolorbox}
\newtcolorbox{deletedcode}{enhanced, sharp corners=south,
        colback=red!5, colframe=red!30, boxrule=0pt,
        borderline west={2pt}{0pt}{red},
        left=6pt, right=2pt, top=2pt, bottom=2pt}
\newtcolorbox{addedcode}{enhanced, sharp corners=south,
        colback=green!5, colframe=green!30, boxrule=0pt,
        borderline west={2pt}{0pt}{green},
        left=2pt, right=2pt, top=2pt, bottom=2pt}

\usepackage{listings}
\usepackage{xcolor} 
\definecolor{LightGray}{gray}{0.9}
\usepackage{xspace}
\usepackage{subfigure}
\usepackage[hidelinks]{hyperref}
\usepackage{orcidlink}
\usepackage[english]{babel}
\usepackage[numbers]{natbib}

\newcommand{\EnergyTrackr}{\textsc{Ener\-gy\-Tra\-ckr}\xspace}

\begin{document}

\title{Systematic Detection of Energy Regression and Corresponding Code Patterns in Java Projects}

\author{Fran\c{c}ois Bechet, NADI, University of Namur, Namur, Belgium
\and J\'er\^ome Maquoi \orcidlink{0009-0006-3576-0757}, NADI, University of Namur, Namur, Belgium
\and Lu\'is Cruz \orcidlink{0000-0002-1615-355X}, Delft University of Technology, Delft, Netherlands
\and Beno\^it Vanderose \orcidlink{0000-0001-9752-0085}, NADI, University of Namur, Namur, Belgium
\and Xavier Devroey \orcidlink{0000-0002-0831-7606}, NADI, University of Namur, Namur, Belgium
}

\date{April 2026}

\maketitle

\begin{abstract}
Green software engineering is emerging as a crucial response to information technology's rising energy impact, especially in continuous development.
However, there remain challenges in devising automated methods for identifying energy regressions across commits and their associated code change patterns.
In particular, little effort has been put into automatically detecting regressions at the commit level by identifying statistically significant changes in energy consumption.
In this paper, we introduce \EnergyTrackr, an approach designed to detect energy regressions across multiple commits that can then be used to identify code anti-patterns potentially contributing to the increase of software energy consumption over time.
We describe our empirical evaluation, including repository mining and source code analysis, made on 3,232 commits from three Java projects, and show the approach's ability to identify significant energy changes. We also highlight recurring anti-patterns such as missing early exits or costly dependency upgrades.
We expect \EnergyTrackr to assist developers in accurately monitoring energy regressions and improvements within their projects, identifying code anti-patterns, and helping them optimize their source code to reduce software energy consumption.

\end{abstract}

\section{Introduction}
\label{sec:introduction}

As the integration of computers continues to play an increasingly prominent role in our daily activities, particularly with the advancements in web and machine learning systems, concerns regarding the rising energy demand have become more pronounced. \citet{Freitag2021RealClimateTransformative} indicate that the global carbon emissions attributable to the Information and Communication Technology (ICT) sector currently range from approximately 2.1\% to 3.9\%, and this proportion is projected to grow over time.

Recently, the research community has dedicated significant effort to exploring various facets of software sustainability and green software engineering~\cite{LeGoaer2023EcoCodeSonarQubePlugin, Hao2013EstimatingMobileApplication, Pang2016WhatProgrammersKnow, Bonvoisin2024UnderstandingPerformanceEnergyTradeoffs, Penzenstadler2012SustainabilitySoftwareEngineering, Pereira2017EnergyEfficiencyProgramming, Poy2025ImpactEnergyConsumption, Simon2025BoaviztAPIBottomUpModel}. Despite these advancements, the field remains in its early development, with numerous challenges yet to be addressed. One such challenge involves identifying commits that may lead to \textit{energy regressions}, an energy consumption increase between two software versions. Detecting these regressions within software projects is complex due to many influencing factors. This step is mandatory to establish a strong empirical basis for identifying common defects causing unnecessarily high energy consumption in the software. We rely on the software's test suite execution to measure its energy consumption. The main idea is that while larger software is expected to consume more energy, significant or abrupt changes may indicate an underlying energy-related issue within the codebase.

One primary challenge involves accurately measuring the energy consumption of software. Several methodologies exist, spanning different levels of granularity, from individual lines of code to entire applications. In this work, we focus on energy measurement at the application level, as it is more portable and introduces less overhead (and, therefore, potential biases) in the measurement. Since we seek to identify energy regressions, we can leverage the commit changes to identify possible underlying code patterns causing them.

Another challenge involves identifying specific code patterns that contribute to increased energy consumption. For instance, since we rely on test execution as a proxy, a change in energy consumption can be caused by changes in the test suite, which is irrelevant for the developer.
\citet{Danglot2024CanWeSpot} introduced the concept of \textit{energy regression testing}. It expands the traditional scope of regression testing (i.e., ensuring that recent changes have not introduced new defects in the pre-existing functionalities) to monitor changes in software's energy consumption, allowing for the detection of variations in energy usage resulting from source code modifications. Complementary to their approach, we seek to provide an overview of the energy consumption throughout the project's history, including potential energy regressions.

This paper introduces \EnergyTrackr, an approach to help minimize the energy consumption associated with software development. 
In a nutshell, it considers the entire (or an arbitrary portion of the entire) commit history, builds each commit individually, and measures its respective energy consumption based on test case executions. \EnergyTrackr provides statistical analysis support with configurable heuristics to identify energy regressions and perform triage, as well as graphical representations of the history to help developers spot problematic code patterns potentially causing an energy regression. A difference in energy consumption between two commits is marked as a regression if the difference in their energy consumption, measured via the execution of their test suites, is statistically significant.

We design and perform an empirical evaluation of \EnergyTrackr to answer the following research questions:
\begin{itemize}
    \item \textbf{RQ1}: To what extent can an automated energy analysis pipeline reliably detect statistically significant energy regressions across commits in real-world Java projects, based on the execution of the test cases?
    \item \textbf{RQ2}: Which recurring change patterns appear in commits flagged by the pipeline as energy regressions?
\end{itemize}
Our analysis of the complete history of three Java projects, corresponding to 3,232 commits,  indicates that \EnergyTrackr can effectively detect significant energy regressions in multiple commits. Additionally, we identified several recurring code modification patterns through manual analysis. Such patterns can serve as a basis for designing an effective energy-aware linter for static code analysis in the future. 

Our contribution includes the open-source implementation of \EnergyTrackr \cite{energytrackr} and the replication package containing the data produced during our evaluation \cite{energytrackr}.

The remainder of this paper is structured as follows: \ref{sec:background} presents the background and related work. The \EnergyTrackr approach is presented in Section \ref{sec:conceptual-solution}, with its implementation described in Section \ref{sec:implementation}. Section \ref{sec:experimental-design} describes our evaluation protocol, with the results presented in Section \ref{sec:results}. Sections \ref{sec:threats-to-validity} and \ref{sec:discussion} discuss the results, current limitations, and new leads for future work. And Section \ref{sec:conclusion} concludes the paper.


\section{Background and related work}
\label{sec:background}

\subsection{Software energy consumption measurement}
\label{sec:approaches-measuring-energy}

There are three main methods to assess the energy consumption of software~\cite{Fahad2019ComparativeStudyMethods}:
\begin{inparaenum}[(i)]
    \item System-level physical measurements utilizing physical power meters.
    \item Measurements obtained through on-chip power sensors.
    \item Energy predictive models.
\end{inparaenum}


\paragraph{Physical power meters}

The first approach involves the usage of physical power meters that offer accurate measurements; however, they do not provide detailed insights at the hardware component level~\cite{Fahad2019ComparativeStudyMethods}. This coarse granularity complicates analysis. The setup is also cumbersome, requiring purchase, calibration, and hardware connection. \citet{Banerjee2014DetectingEnergyBugs} automates energy hotspot detection in Android apps using a physical power meter connected to smartphones. Similarly, \citet{Schubert2012ProfilingSoftwareEnergy} used the \textit{Watts Up? Pro} power meter to evaluate the \texttt{eprof} profiler. \citet{Hindle2014GreenMinerHardwareBaseda} proposed a tool named GreenMiner, which measures the energy consumption of mobile devices, automates application testing processes, and generates detailed reports for developers and researchers. They demonstrated the method's efficacy by showing that it could help gather measurements faster by using parallel and continuous properties and by finding energy regressions in changes related to the user interface. It offers high stability, but the cost of this stability is physical hardware dependency on their in-house hardware setup. Unlike these studies that use external power meters, \EnergyTrackr operates without needing physical hardware power meters and is not restricted to mobile devices.

\paragraph{On-chip power sensors}

The second approach involves the usage of on-chip power sensors provided by manufacturers. Such hardware sensors with software interfaces can measure components like CPUs, GPUs, disks, and RAMs. For example, Intel's Running Average Power Limit (RAPL) provides CPU electricity consumption data to the operating system (OS)~\cite{Raffin2025DissectingSoftwarebasedMeasurement}. RAPL is recognized as one of the most accurate tools for measuring processor energy consumption~\cite{SpencerDesrochers2016ValidationDRAMRAPL, Khan2018RAPLActionExperiences}, which contributes to its widespread use in research studies focused on CPU power measurement~\cite{Liu2015DataOrientedCharacterizationApplicationLevel, Jay2023ExperimentalComparisonSoftwarebased, Schuler2022MANAiIntelliJPlugin}. In software energy consumption, the literature primarily emphasizes CPU measurement~\cite{Pathania2025CalculatingSoftwaresEnergy}, which helps explain the widespread use of RAPL among researchers. For example, \citet{Ournani2021EvaluatingEnergyConsumption} use it to directly compare the energy consumption of various Java I/O libraries and methods. Additionally, RAPL can be integrated with specialized tools 
such as PowerAPI~\cite{Fieni2024PowerAPIPythonFramework}, JoularJX~\cite{Noureddine2022PowerJoularJoularJXMultiPlatform}, TLPC-sensor~\cite{Danglot2024CanWeSpot} or the \texttt{perf} Linux utility.

PowerAPI is a tool that offers real-time insights into an application's power consumption~\cite{Fieni2024PowerAPIPythonFramework}. It is based on machine learning methods to estimate the energy consumption of applications and software based on raw performance metrics, and it automatically calibrates itself as needed. In contrast, \EnergyTrackr operates without machine learning, focusing on tracking changes in energy consumption across different Git commits.

JoularJX is designed for monitoring power usage at the source code level of Java projects and handles different granularity levels~\cite{Noureddine2022PowerJoularJoularJXMultiPlatform}. However, these strengths also introduce limitations as the tool only focuses on Java, and lower granularity can cause some imprecision, amplifying environmental noise. This limitation is caused by RAPL, which is designed for aggregate measurements, not low-level precision. In contrast, \EnergyTrackr operates at a higher granularity level and does not have these limitations.

TLPC-sensor is a tool that collects energy consumption and performance metrics to facilitate analysis of the relationship between these factors~\cite{Danglot2024CanWeSpot}. The measurements are performed using RAPL in a time-based manner, initiated and terminated through explicit calls to the \(start()\) and \(stop()\) functions. This approach gives the user precise control over what is measured, allowing for measurements at various levels of granularity, such as specific code blocks. While this instrumentation capability offers flexibility, it also introduces increased complexity to the tool. In contrast, \EnergyTrackr does not require any source code instrumentation to perform its energy measurements.

\texttt{Perf} is a widely recognized Linux utility for collecting energy measurement data. One key advantage of utilizing \texttt{perf} over manual reading of RAPL driver files is its higher sampling rate and lower overhead, due to its integration within the Linux kernel. Due to its ability to deliver high-quality energy measurements, this study employed this tool to gather relevant data. In a previous study, the authors compared \texttt{perf} with alternative RAPL-based tools, concluding that these tools produce highly consistent results~\cite{Jay2023ExperimentalComparisonSoftwarebased}. The findings indicate that \texttt{perf} is sufficiently reliable when complemented with additional techniques such as repeated measurements.

\paragraph{Energy predictive models}

The third approach uses theoretical models to estimate consumption without direct measurements. For example, \citet{Acar2016ImpactSourceCode} proposed a model to estimate the CPU, memory, and disk power consumption during the execution of the application using established mathematical formulas. These models offer the benefit of eliminating the need for additional infrastructure or time-consuming measurement processes, while enabling detailed, component-level analysis of energy consumption during application execution. However, existing research indicates that their overall accuracy may be limited~\cite{McCullough2011EvaluatingEffectivenessModelbased, Obrien2017SurveyPowerEnergy, Shahid2017AdditivitySelectionCriterion}. Unlike these predictive models, \EnergyTrackr will use the tool \texttt{perf} to measure CPU energy consumption via the RAPL interface accurately.

\subsection{Challenges in energy assessment}
\label{sec:challenges-energy-assessment}

\paragraph{Granularity}
In the context of power measurement, various levels of granularity can be considered, ranging from the analysis of individual lines of code to the overall energy consumption of the entire device. \citet{Harman2015AchievementsOpenProblems} describe three distinct levels of granularity:
\begin{inparaenum}[(i)]
    \item Fine-grained level, which evaluates the contribution of each line of code to the overall energy consumption. For example, \citet{Li2013CalculatingSourceLine} describe an approach that provides developers with source line level energy information by combining hardware-based energy measurements with program analysis and statistical modeling techniques; \citet{Maquoi2025EnergyCodesumptionLeveraging} identify energy-intensive source code constructs by analyzing the correlation between energy consumption and individual lines of Java code, and suggest a link between energy consumption and the number of objects' attributes created within that code.
    \item Mid-grained level focuses on the energy utilized by a specific block, method, or procedure within the program. For example, \citet{Hahnel2012MeasuringEnergyConsumption} use RAPL to measure the power consumption of short-running paths, such as the energy cost of a single function within an application or a system call of the underlying OS; \citet{Schubert2012ProfilingSoftwareEnergy} present eprof, a software energy profiler that can identify the energy used by individual functions. \EnergyTrackr lies within the mid-grained granularity level, as the objective is to find energy regressions across commits.
    \item Coarse-grained level measures the total energy consumption associated with executing the program or process over a designated period. For example, \citet{Mancebo2021DoesMaintainabilityRelate} present the FEETINGS framework, designed to measure, analyze, and visualize the energy consumption of software. This framework investigates possible correlations between software maintainability metrics, such as the total lines of code, and the corresponding energy consumption.
\end{inparaenum}

\paragraph{Reproducibility}

Energy or power measurements in a scientific study are far from trivial. Even when performed on the same hardware, these measurements can be influenced by various factors such as component temperatures (CPU, GPU, memory), ambient room temperature, background system processes, and thread scheduling~\cite{Cruz2021GreenSoftwareEngineering}. This inherent variability presents challenges when evaluating the energy efficiency of specific software versions. For instance, relying on a single measurement per test may lead to inconsistent results, as repeated trials can yield varying outcomes.

\citet{Cruz2021GreenSoftwareEngineering} proposed the following method, which has been widely adopted by the green software engineering community~\cite{Roque2025UnveilingEnergyVampires, Xiao2025EffectivenessMicroservicesTactics, Georgiou2022GreenAIDeep} and is also used for the development of \EnergyTrackr:
\begin{compactitem}
    \item[\textbf{Zen mode}:] Minimize the computer environment impact, for example, by closing as many background processes as possible or by turning off auto-brightness mode on the devices.
    \item[\textbf{Freeze your settings}:]  Save the current state of the machine and reuse that same setting during the entire testing process.
    \item[\textbf{Warm-up}:] A CPU performs more efficiently when its temperature stabilizes after idle. A calibration or warm-up phase is essential before measurements to reach the target temperature, preventing fluctuations from affecting initial results and ensuring accuracy, as noted by \citet{Ournani2021EvaluatingImpactJava}.
    \item[\textbf{Repeat}:] To address measurement instability, perform multiple runs~\cite{Arcuri2014HitchhikersGuideStatistical} and use the median as the final value. \citet{Cruz2021GreenSoftwareEngineering} recommends 30 measurements, but practical constraints often lead to fewer, such as 10 repetitions  \cite{Bangash2017MethodologyRelatingSoftware}. Some studies, like \citet{Danglot2024CanWeSpot}, use up to 100 repetitions.
    \item[\textbf{Rest}:] Temperature influences CPU energy efficiency, causing a tail effect where the end of one measurement affects the next. To mitigate this, \citet{Cruz2021GreenSoftwareEngineering} recommends resting the machine between tests—typically around one minute—to allow temperatures to stabilize. While it does prolong the measurement process, this duration balances accuracy with evaluation cost.
    \item[\textbf{Shuffle}:] Repeating measurements for the same commit can be affected by environmental factors like background processes. Creating batches of tasks, e.g., measuring $n$ commits $k$ times results in $n \times k$ tasks, can help mitigate this variability. Randomizing the order of these tasks can further reduce environmental impact and improve the reliability of measurement results.
    \item[\textbf{Keep it cool}:] A CPU getting too hot can start throttling and impacting the measurements. Therefore, ensuring the CPU stays cool during the process is very important. Collecting CPU temperature data during the measurements can also be a good way to ensure it does not impact the measurements.
    \item[\textbf{Automate executions}:] Reproducibility is impossible without automating the whole measurement process. Writing a script for the measurement execution is a mandatory step.
\end{compactitem}

\subsection{Energy regressions}
\label{sec:previous-research}

Energy regression testing is defined by \citet{Danglot2024CanWeSpot} as detecting energy consumption drifts in software systems induced by changes observed at the source code level, a.k.a energy regressions. Their study emphasizes the challenge of effectively notifying developers about increased energy consumption in applications during CI processes. To do so, they gather energy measurements via the execution of the tests of the analyzed software at specific commits. Only the tests that cover the source code changes introduced by each commit are executed to optimize testing efficiency. The study suggests that developers' tests can identify energy regressions resulting from code changes, although the effectiveness of this approach varies depending on the configuration. This study is a foundation for further research, as additional work remains. Conversely, \EnergyTrackr does not utilize Continuous Integration (CI) for energy regression detection, as RAPL often does not have access to such development environments. Instead, it employs \texttt{perf} to measure the energy consumption associated with code commits accurately. In addition to this methodology, \EnergyTrackr offers a comprehensive overview of the project's energy consumption history, including detection of any potential regressions in energy efficiency.

Other approaches for the identification of energy regressions exist. For instance, \citet{Poy2025ImpactEnergyConsumption} conducted a systematic mapping examining how design patterns, code smells, and refactoring techniques influence software energy consumption. Their findings indicate that the usage of design patterns does not always positively impact software energy consumption. For example, positive impacts are observed in some cases, such as with the Flyweight and Interpreter patterns. In contrast, other patterns, including the Chain of Responsibility, Decorator, or Iterator, were associated with increased energy consumption. The results were inconclusive for specific patterns like Abstract Factory, Adapter, and Bridge due to conflicting findings across primary studies. 

Concerning the code smells, the topic is complex. \citet{Poy2025ImpactEnergyConsumption} suggests that removing nearly all code smells in their primary studies enhances software energy efficiency. However, \citet{ConnollyBree2025HowSoftwareDesign} conducted a systematic literature review examining the impact of various design elements, including code smells, on software energy consumption. Their findings indicate that removing code smells does not always reduce energy usage. This may be because identifying and addressing code smells can be more challenging than implementing design patterns. The relationship between code smells and refactoring techniques can also introduce potential confusion. For instance, the Move Method refactoring technique can decrease software energy when fixing the Feature Envy code smell. Conversely, if misapplied, the Move Method can inadvertently introduce this code smell and increase energy consumption. Further research is needed to understand better instances where refactoring efforts may cause energy regressions related to code smells. In contrast, \EnergyTrackr does not analyze individual design elements or evaluate the effects of specific design patterns against unpatterned source code. Instead, it is tailored to identify energy regressions by studying the evolution of code commits over time and can serve as a basis for future research on those topics.

\section{\EnergyTrackr approach}
\label{sec:conceptual-solution}

\label{sec:conceptual-architecture}

In a nutshell, the core idea of \EnergyTrackr is as follows:
\begin{inparaenum}[(i)]
  \item running over each commit of a project under test to measure its energy consumption based on the execution of the test cases;
  \item sorting the result, as \EnergyTrackr runs tasks in a random order to counter environment effects on energy measurements;
  and \item building a report, implemented as a web page (allowing for easy sharing), containing the analysis of the results and flagging potential energy regression commits.
\end{inparaenum}

\subsection{Preparation phase}
\label{sec:preparation-phase}

This phase stabilizes the Linux machine for repeatable and stable energy measurements by turning off power-saving features and pinning frequencies (\textbf{Zen mode}) via a configuration file. All modifications are reversible, as the device parameters are copied rather than directly altered, allowing users to easily switch between defined configurations of \EnergyTrackr (\textbf{Freeze your settings}).

\paragraph{Stability verification}

A stability verification step is incorporated before conducting the measurements (\textbf{Warm-up}). This step aims to confirm that the system is in a stable and consistent energy state. This step can also be executed in the following stages if the environment is unstable. During this process, the modified z-score~\cite{Iglewicz1993Volume16How} statistical method will be employed to identify anomalous data points from energy values (directly obtained from the RAPL interface in our implementation). The modified z-score $z$ for a sample $x$ is computed as:
  \[
    z = \frac{0.6745 \cdot (x - \tilde{x})}{\text{MAD}}
  \]
where $\tilde{x}$ is the median of the samples, $\text{MAD} = \text{median} \left( \left| x_i - \tilde{x} \right| \right)$ is the Median Absolute Deviation, and $x_i$ are the individual sample values.

This stability verification process begins by collecting five initial warm-up samples to establish a baseline. Subsequently, measurements are taken from 30 additional samples, each individually compared to the established baseline. If any measurement's modified z-score exceeds the predefined threshold of 3.5, the system is flagged as potentially unstable, and the process is halted. The modified z-score is utilized instead of standard deviation due to its robustness in handling non-normal data distributions and its reliability with small or skewed datasets. Energy differences are analyzed rather than absolute energy values, as the latter can increase over time (e.g., the RAPL interface reporting cumulative energy consumption since system startup). In this context, the rate of change between samples is of primary interest for stability assessment.

\subsection{Measurement phase}
\label{sec:measurement-phase}



The selected approach for energy measurements involved wrapping the entire test suite execution (with \texttt{perf} in our implementation) rather than instrumenting individual tests for greater precision. This method offers benefits such as reduced overhead and improved measurement accuracy. Specifically, the increased sampling duration enables more reliable statistical analysis by mitigating the impact of environmental variability, such as background processes.

First, \EnergyTrackr clones the target repository and collects the relevant commit history for analysis. An initial preprocessing stage is conducted, which involves filtering out unnecessary commits. 
Our implementation also includes a check that \texttt{perf} can be used without root permissions. 
Commits are then organized into batches, which are randomized to counter environmental effects on measurement quality (\textbf{Shuffle}). After that, all commits within each batch are built. The energy consumption measurement is done based on the test suite execution. The test execution command can be configured to accommodate different build systems and test environments (e.g., \texttt{mvn test} for Java projects). The pipeline can be configured to run the same commit a specified number of times (\textbf{Repeat}), with the default setting of 30 repetitions as recommended by \citet{Arcuri2014HitchhikersGuideStatistical}. In our implementation, a custom bash script was designed to set up the machine on which \EnergyTrackr will run (\textbf{Automate executions}).

The current CPU temperature is monitored to ensure it remains within safe operational limits (\textbf{Keep it cool}). If temperatures are too high, the process pauses until cooling conditions are achieved (\textbf{Rest}). Energy consumption measurements are then collected for each commit in the batch and compiled for analysis.


\subsection{Regression detection phase}
\label{sec:regression-detection-phase}

Given the randomized execution of the measurements, the data needs sorting before analysis. In our implementation, this step produced CSV files. The data for the different commits are grouped by commit and sorted in descending order by date before being preprocessed.

\paragraph{Data preprocessing}


The first stage checks, for each commit, whether the data follows a normal distribution using the Shapiro-Wilk test \cite{ShapiroWilk1965} for the different energy measures for that commit. It is important to note that commits with non-normal distributions can still be classified as significant changes. This is because distribution normality is not part of the significance classification itself but serves as an additional measurement quality indicator.

Second, following best practices \cite{Cruz2021GreenSoftwareEngineering}, we filter out outliers. Specifically, this second stage detects transient outlier commits ---drops in sustained energy that indicate potential outliers rather than real regressions--- by using the interquartile range (IQR) statistic, with a Tukey multiplier of 1.5~\cite{Tukey1949ComparingIndividualMeans}. The stage allows some configuration variables as input, like the rolling window range of commits used to detect outliers, the Tukey multiplier for IQR filtering, the maximum number of transient outlier commits, and the aggregation method used to aggregate the multiple measurements done for every commit.

\paragraph{Change detection}

This third stage identifies significant changes or regressions in the energy data and sorts them into five levels. A commit is first tested for level 1, then for level 2, etc, until level 5. The highest level the change satisfies determines its classification, meaning a change can satisfy multiple levels or fail some intermediate ones. As those settings can be project-dependent, they can be fully configured. We report here the levels with the default settings used in our implementation (and evaluation):

\begin{compactitem}\label{list:changes-levels}
  \item[\textbf{Level 1}:] Performs a parametric Welch's unequal variances t-test \cite{Welch1947} to assess statistical significance with a threshold $\alpha = 0.05$. If the commit passes this test, then it is classified as \textit{level 1} and considered a regression/improvement.
        
  \item[\textbf{Level 2}:] Calculates Cohen's $d$ \cite{Cohen1988} to measure effect size. It uses the following thresholds to classify the change:
        \begin{inparaenum}
          \item[negligible] for \( |d| \leq 0.2 \);
          \item[small] for \( 0.2 < |d| \leq 0.5 \);
          \item[medium] for \( 0.5 < |d| \leq 0.8 \);
          \item[large] for \( |d| > 0.8 \). 
        \end{inparaenum}
        If the commit is at least in the small category ($0.2<|d|$), then it is classified as \textit{level 2}.

  \item[\textbf{Level 3}:] Computes the relative change \( \Delta\%\) on the median energy values. Let \( m_b \) be the median of the baseline sample and \( m_t \) be the median of the test sample, the relative change for a pair of commits is computed like:
        \[\text{Percentage Change} = \left| \frac{m_t - m_b}{m_b} \right|\]
        It uses the following thresholds to classify the change:
        \begin{inparaenum}
          \item[minor] for \( \Delta\% < 5.0\% \);
          \item[moderate] for \( 5.0\% \leq \Delta\% < 10.0\% \);
          and \item[major] for \( \Delta\% \geq 10.0\% \).
        \end{inparaenum}
        If the commit is at least moderate, then it is classified as \textit{level 3}.

  \item[\textbf{Level 4}:] Computes the practical significance \( \Delta J\), i.e., how strong the regression is relative to the baseline. Let \( m_b \) be the median of the baseline sample and \( m_t \) be the median of the test sample, the relative change for a pair of commits is computed like:
        \[
          \Delta J = m_t - m_b
        \]
        It uses the following thresholds to classify the change:
        \begin{inparaenum}
          \item[info] for \( \Delta J < 5.0\% \times m_b \);
          \item[warning] for \( 5.0\% \times m_b \leq \Delta J < 10.0\% \times m_b \);
          \item[critical] for \( \Delta J \geq 10.0\% \times m_b \).
        \end{inparaenum}
        If the commit is at least in the warning category, then it is classified as \textit{level 4}. Improvements will consistently be categorized under the "info" classification and, as a result, will not attain level 4. This is because they will always exhibit a negative \( \Delta J \) and consequently remain below 5.0\% of \(m_b\).

  \item[\textbf{Level 5}:] When context tags from \EnergyTrackr's configuration are present in the commit message, then the commit will be flagged to \textit{level 5}. For example, if the commit message contains the word `\textit{performance}' and that the tag `\textit{performance}' is in the tag list.
\end{compactitem}
\citet{Cruz2021GreenSoftwareEngineering} and \citet{Danglot2024CanWeSpot} inspired the design of levels 1 to 3. The additional levels were designed to provide a more exhaustive and comprehensive classification of changes and were developed specifically for \EnergyTrackr. This combined approach ensures that only relevant, meaningful changes are reported, filtering out random fluctuations while highlighting regressions or improvements developers should address in priority.

\subsection{Data visualization phase}
\label{sec:report-generation-phase}


To enable developers to analyze the results and identify energy regressions based on our classification from levels 1 to 5, \EnergyTrackr proposes different energy consumption data visualizations.

\paragraph{Data visualization}

\begin{figure}[t]
    \centering
    \subfigure[Evolution plot showing the commit \href{https://github.com/jhy/jsoup/commit/cc2363e4501e086b6ba628ececb7716cfad87796}{\texttt{cc2363e}} and nearby commits from \texttt{JSoup}.\label{fig:evolution-plot-jsoup}]{\includegraphics[width=1\textwidth]{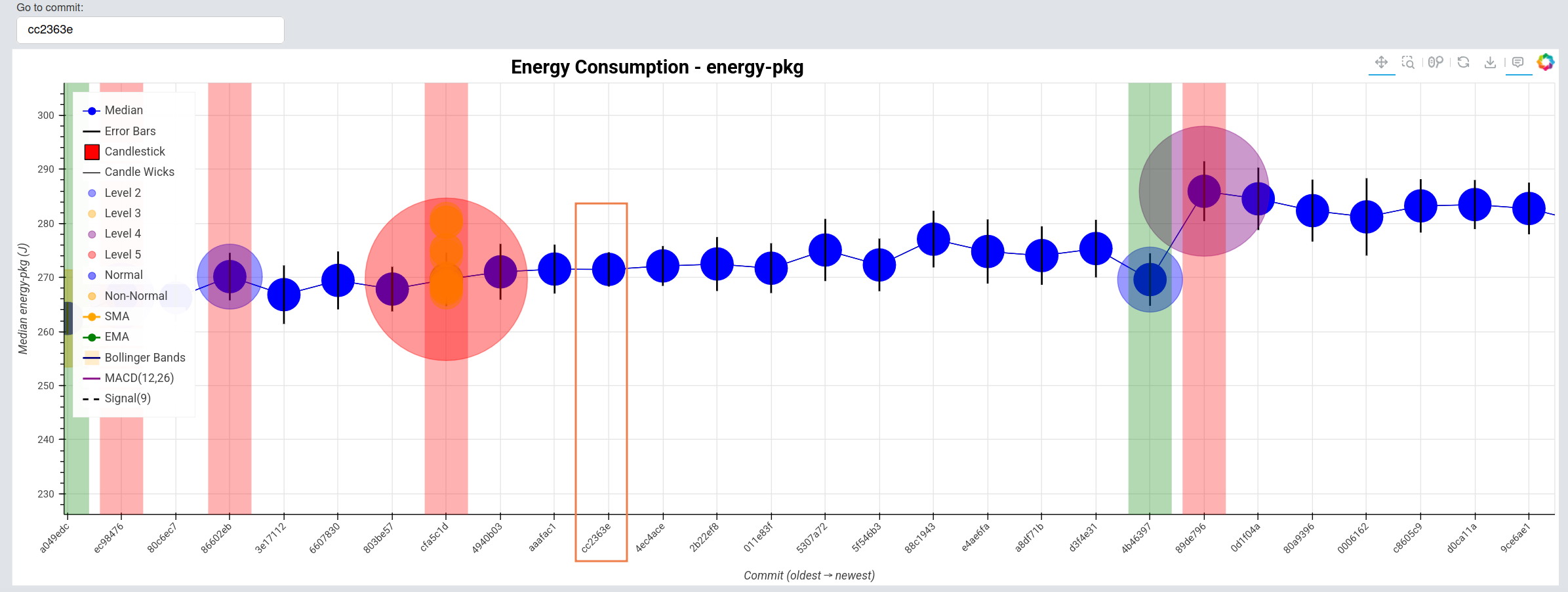}}
    \subfigure[Cusum plot from \texttt{JSoup}'s history.\label{fig:cusum-plot-jsoup}]{\includegraphics[width=0.48\textwidth]{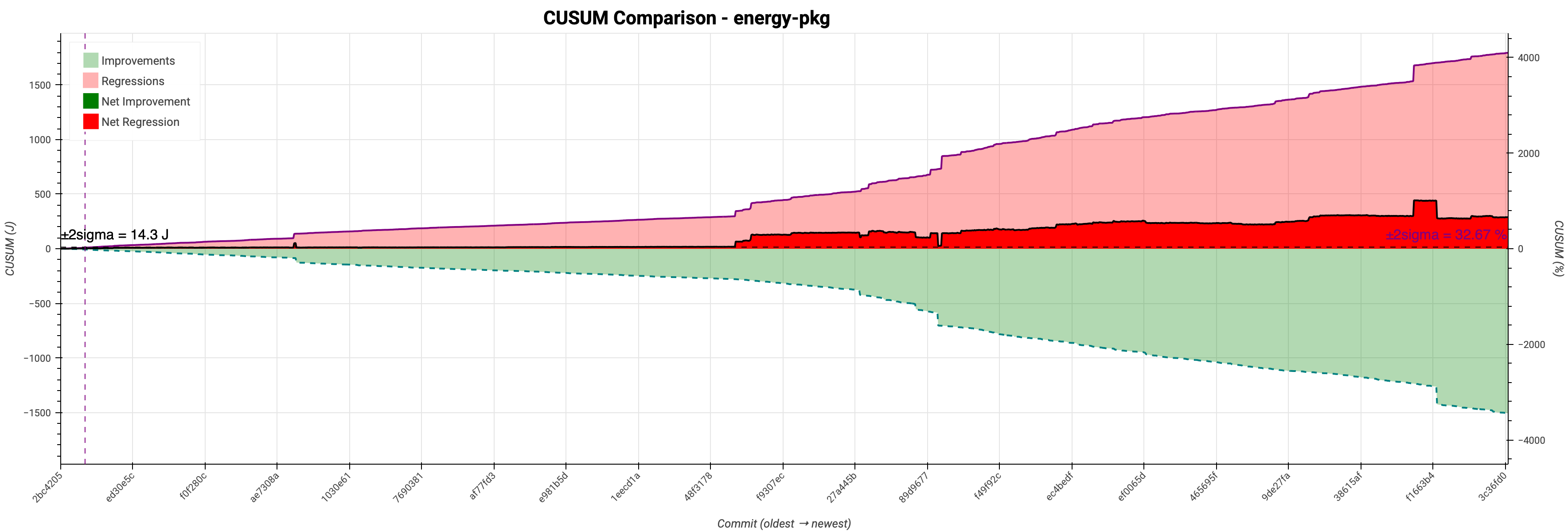}}
    \subfigure[Change points detection plot from \texttt{JSoup}'s history.\label{fig:change-points-jsoup}]{\includegraphics[width=0.48\textwidth]{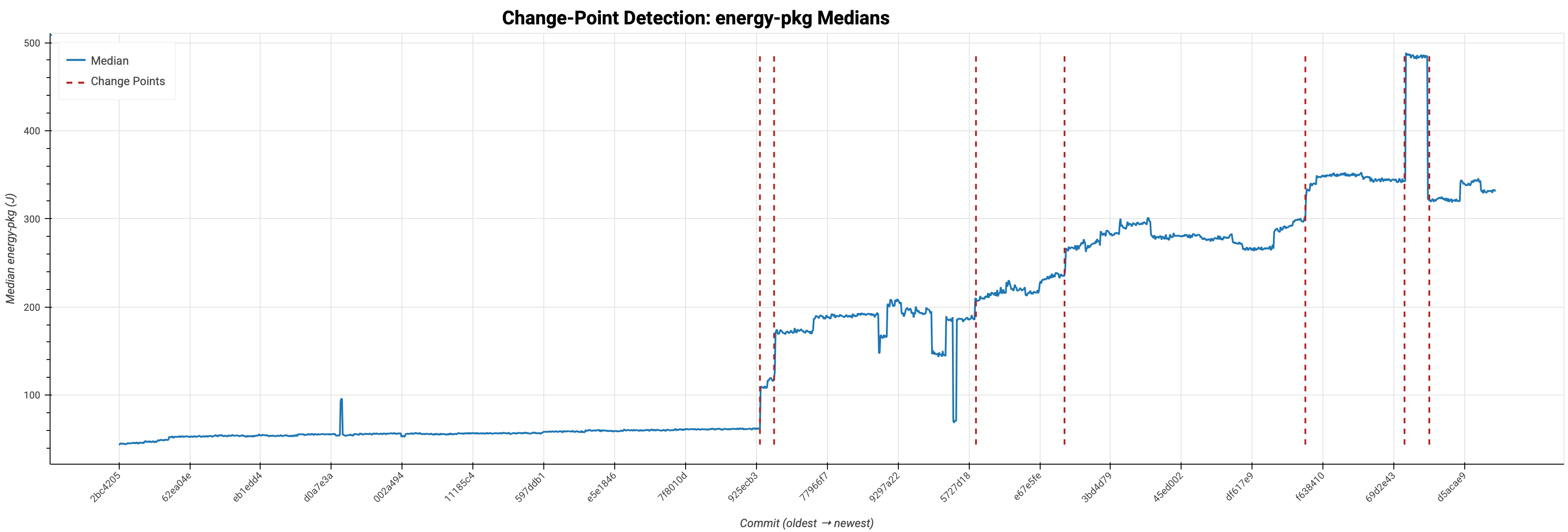}}
    \caption{Excerpt of the different plots generated for \texttt{JSoup} energy measurement.}
    \label{fig:plotsjsoup}
\end{figure}

This stage creates interactive plots to enable data visualization. \EnergyTrackr's implementation supports the following plots:
\begin{compactitem}
    \item[\textbf{Evolution plot}] plot shows the history of the commit range. For instance, Figure \ref{fig:evolution-plot-jsoup} presents the evolution of energy consumption for a portion of JSoup's commit history. Each blue circle indicates a commit. The red and green vertical bars represent commits with a level of at least 2, whether they signify an energy regression (red) or an improvement (green). Some commits are highlighted with a colored circle encircling them. These circles denote the level of each commit, starting from level 2, with size and color corresponding to the specific level: a blue circle for level 2, an orange circle for level 3, a purple circle for level 4, and a red circle for level 5. For example, to the right of the Figure \ref{fig:evolution-plot-jsoup}, a level 4 regression is visible, indicating that this commit has been classified as "warning," with a negative impact estimated between 5.0\% and 10.0\% compared to the previous commit.

    \item[\textbf{Boxplots and violin plots}] allow for comparing two commits in terms of the energy consumption distribution across the repeated executions of their respective test suites~\cite{Hintze1998ViolinPlotsBox, Dutoit2012GraphicalExploratoryData}.

    \item[\textbf{Quantile-Quantile plot}] compares the distributions of the energy consumption of two commits~\cite{Wilk1968ProbabilityPlottingMethods}.

    \item[\textbf{Bootstrap plot}] is an interactive histogram that helps confirm if the difference between two commits is statistically significant~\cite{Efron1993IntroductionBootstrap}.

    \item[\textbf{Cusum plot}] shows the cumulative sum of the energy of improvements and regressions. As shown in Figure \ref{fig:cusum-plot-jsoup}, it gives an overview of the macro scale of a project to assess the global energy impact.

    \item[\textbf{Change point detection plot}] detects change points in the median energy data. As shown in Figure \ref{fig:change-points-jsoup}, it gives another overview of a project's abrupt changes in energy consumption.
\end{compactitem}


\paragraph{Report generation}

The final stage of \EnergyTrackr is to gather the different plots and generate a report with additional information, such as the level's configuration with the various thresholds, and a table listing the different commits with their metadata and statistics. This report can serve as a basis for developers to identify and debug energy regressions.


\section{Open source implementation}
\label{sec:implementation}

\EnergyTrackr is implemented as an open-source tool, available on GitHub \cite{energytrackr}. 
It follows a pipes and filters design pattern that offers great modularity.
Each stage described in Section \ref{sec:conceptual-solution} is implemented as a separate module, allowing for easier testing and future extensibility.
It is possible to add a new stage to \EnergyTrackr by defining a new module, or to modify the current behavior by removing or altering a stage in the configuration file. 
This choice of modular design with proper separation between configuration, energy measurement, plotting, and shared utilities was deliberate to ensure:
\begin{inparaenum}[(i)]
    \item cross-language capabilities in both the analysis of different languages projects (currently, only Java projects are supported), but also allowing interaction with other measurement tools written in different languages through an easier API interaction;
    \item a better reproducibility thanks to a centralized configuration file, which is an important quality factor when taking energy measurements;
    and \item an easier integration into existing developers' workflow.
\end{inparaenum}

The implementation includes two main configuration files to configure \EnergyTrackr for projects running in different environments. The initial configuration file enables developers to define variables related to the measurement pipeline, such as the repository URL, the number of commits to iterate over, or the granularity level (e.g., commit, branch, or tag). The second configuration file allows users to define which data transformations to apply and how they are configured, specify which plots to generate along with their settings, and determine the page components to include and customize for building the final report.

To facilitate seamless integration with other tools and ensure comprehensive pipeline management, \EnergyTrackr's execution is orchestrated using a Continuous Integration (CI) system. This design enhances compatibility with existing developer workflows and enables complete automation of processes.

\section{Evaluation setup}
\label{sec:experimental-design}


The evaluation will aim to validate our approach for energy regression detection (\textbf{RQ1}) and identify regression commits that offer insights into their underlying causes and the recurrence of identified patterns (\textbf{RQ2}). 
This evaluation is both quantitative and qualitative. The quantitative part (\textbf{RQ1}) involves executing \EnergyTrackr as described in Section \ref{sec:conceptual-solution} across a selection of real-world GitHub projects with strong test suites, measuring energy consumption, performing the data analysis as specified in Section \ref{sec:regression-detection-phase}, and building the reports. %
The qualitative part (\textbf{RQ2}) consists of manually inspecting the report generated by the pipeline to identify common code-level patterns or changes responsible for energy regressions. 

\subsection{Repository selection}
\label{sec:repository-selection}

To answer our RQs, we selected three Java projects, based on the following inclusion criteria:
\begin{inparaenum}[(i)]
    \item a \textit{long commit history} with at least 500 commits to allow a rich comparison; 
    \item a \textit{good test suite} that can be executed reliably over the commit history;
    \item \textit{open source} projects recommended in the \textit{Awesome-Java} list of open source projects;\footnote{Available at \url{https://github.com/akullpp/awesome-java}}
    \item relying on a \textit{dependency management system} like Maven, which is important to be able to run tests over a long history;
    and \item the project should use the \textit{Git version control system}.
\end{inparaenum}

\begin{table}[t]
    \caption{List of selected projects with their characteristics.}
    \label{tab:project_overview}
    \centering
    \begin{scriptsize}
        \resizebox{\textwidth}{!}{\begin{tabular}{l l l l}
\toprule
                                   & \textbf{Jsoup}       & \textbf{univocity-parsers} & \textbf{fastexcel}              \\
\midrule

\textbf{Commit SHA}    
    & \href{https://github.com/jhy/jsoup/commit/01b3900107114f441860ee3d03aba03caef42804}{01b3900}        
    & \href{https://github.com/uniVocity/univocity-parsers/commit/7e7d1b3c0a3dceaed4a8413875eb1500f2a028ec}{7e7d1b3}
    & \href{https://github.com/dhatim/fastexcel/commit/66a9dcb6022286d99f6d7252e161e4e58baa7bd9}{66a9dcb} \\
\textbf{JDK version}                           
    & 24.0.1     
    & 1.6
    & 1.8               \\ 
\textbf{Lines of Code (LOC)}
    & 9450    
    & 9700
    & 2222             \\ 
\textbf{Total / Failed / Ignored tests}                        
    & 1490 / 0 / 46
    & 1106 / 2 / 0
    & 81 / 0 / 0                \\
\textbf{Line Coverage}                     
    & 88\% (8406 / 9450)              
    & 77\% (7519 / 9700)
    & 87\% (1950 / 2222)                \\
\textbf{Branch Coverage}                
    & 83\% (4170 / 4976)                 
    & 74\% (4220 / 5651)
    & 70\% (699 / 997)                \\
\bottomrule
\end{tabular}}
    \end{scriptsize}
\end{table}

The selected projects are \texttt{Jsoup}, \texttt{univocity-parsers}, and \texttt{fastexcel}, listed in Table \ref{tab:project_overview} with their characteristics, derived from the most recent commit examined for each project. \texttt{Jsoup} is a Java library designed to parse, extract, and manipulate HTML documents. \texttt{univocity-parsers} is a fast and reliable collection of Java parsers. The project has been inactive since Apr 19, 2021. \texttt{fastexcel} is a Java library designed to quickly read and generate Excel files.
The balance between analysis quality and feasibility can explain the limited number of projects. Indeed, the energy consumption measurement for each commit is a long process with repeated steps to ensure the quality of the results.

\subsection{Hardware and software configuration}
\label{sec:hardware-and-software-environment}

\begin{table}[t]
    \caption{Machines specifications}
    \label{tab:hardwares-specs}
    \centering
    \begin{scriptsize}
        \begin{tabular}{lp{40mm}p{40mm}}
\toprule
        & \textbf{First machine}      & \textbf{Second machine}              \\
\midrule
\textbf{Processor}    
    & Intel Core i7-14700K (28) @ 4.30\,GHz
    & AMD Ryzen 9 7900X (24) @ 4.70\,GHz \\
\textbf{Motherboard}                           
    & Gigabyte Z790 Aorus Elite AX
    & MSI MPG B650 Tomahawk \\ 
\textbf{Cooling system}
    & Water cooling -- Thermalright Frozen Magic 360 Scenic V2 
    & Thermalright Peerless Assassin 120 SE \\ 
\textbf{PSU}                    
    & NZXT C850 -- 80 Plus Gold
    & Corsair HX1000i -- 80 Plus Platinum \\
\bottomrule
\end{tabular}
    \end{scriptsize}
\end{table}

We used two machines to provide the energy results, each equipped with high-end CPUs, proper cooling systems, and certified Power Supply Units (PSUs) to maximize stability. The specifications of the two machines are presented in Table \ref{tab:hardwares-specs}.
Using one machine would have provided more consistent results, but would have slowed down the measurements, which already take considerable time to gather. For example, for a batch of 100 commits for \texttt{Jsoup}, building takes less than 1 minute on the first machine, thanks to multiprocessing. However, test measurements take more time. For \texttt{Jsoup}, it takes on average 3 seconds per commit, so the total time is: $ 
T_{\text{total}} = T_{\text{build}} \times N_{\text{measures}} \times N_{\text{commits}}
= 3\,\text{seconds} \times 15 \times 2200
= 99\,000\,\text{seconds} \approx 1\,\text{day and } 3.5\,\text{hours}$.
We used the first machine to collect data for \texttt{Jsoup} and the second machine for \texttt{univocity-parsers} and \texttt{fastexcel}. In total, it took us 3 days to execute the evaluation.

It is important to note that, by default, 30 measurements should be made, as recommended by \citet{Arcuri2014HitchhikersGuideStatistical}. However, after comparing the same analysis with 15 and 30 measurements, as they were very close, we decided to drop to 15 to speed up data collection for \texttt{JSoup} and \texttt{fastexcel}, as our time to execute the projects and collect the data on the machines described in Table \ref{tab:hardwares-specs} was limited. 
\section{Evaluation Results}
\label{sec:results}

The \texttt{JSoup} project was tested on the first machine, while the \texttt{univocity-parsers} project was tested on the second machine. The \texttt{fastexcel} project was tested on the second machine. However, some interesting commits were excluded during the analysis step due to a non-normal distribution. A second run of the project with more repetitions, instead of the 15 used, could potentially allow looking at those interesting regressions. However, this was not possible in our case due to the machines not being available. Those non-normally distributed commits are excluded from the analysis.

\subsection{Energy regression detection (RQ1)}

\begin{table}[t]
    \caption{Results summary}
    \label{tab:case-studies-summary}
    \centering
    \begin{scriptsize}
        \begin{tabular}{ll}
\toprule
\multicolumn{2}{c}{\textbf{Jsoup}} \\
\midrule
\textbf{Total commits} & 1944 \\
\textbf{Commit range} &
\href{https://github.com/jhy/jsoup/commit/2bc420589478b3cb01398cd9eb233be25b73b7c0}{2bc4205}
--- 
\href{https://github.com/jhy/jsoup/commit/01b3900107114f441860ee3d03aba03caef42804}{01b3900} \\
\textbf{Commit dates} & (2011-07-02) --- (2025-04-03) \\
\textbf{Significant changes} & 216 (124 regressions) \\
\midrule
\textbf{Mean energy} & 190.43 J \\
\textbf{Median energy} & 190.86 J \\
\textbf{Std. dev.} & 126.51 J \\
\midrule
\textbf{Normal dist.} & 1801 / 1944 (92.64\%) \\
\textbf{Outliers removed} & 42 \\
\textbf{Relative change $\Delta\%$} &
190 minor, 12 moderate, 14 major \\
\textbf{Practical significance $\Delta J$} &
197 info, 10 warning, 9 critical \\
\textbf{Level 2 / 3 / 4 / 5} &
155 / 33 / 18 / 10 \\
\midrule
\multicolumn{2}{c}{\textbf{univocity-parsers}} \\
\midrule
\textbf{Total commits} & 788 \\
\textbf{Commit range} &
\href{https://github.com/uniVocity/univocity-parsers/commit/e536b75ab58d0500552c6d80f09f6463d70a91d8}{e536b75}
--- 
\href{https://github.com/uniVocity/univocity-parsers/commit/ed3d0d3848d1ab96aba7ea4883d5c208b6f4c5e5}{ed3d0d3} \\
\textbf{Commit dates} & (2014-07-12) --- (2021-04-19) \\
\textbf{Significant changes} & 381 (189 regressions) \\
\midrule
\textbf{Mean energy} & 127.05 J \\
\textbf{Median energy} & 131.56 J \\
\textbf{Std. dev.} & 9.99 J \\
\midrule
\textbf{Normal dist.} & 691 / 788 (87.69\%) \\
\textbf{Outliers removed} & 25 \\
\textbf{Relative change $\Delta\%$} &
376 minor, 2 moderate, 3 major \\
\textbf{Practical significance $\Delta J$} &
377 info, 2 warning, 2 critical \\
\textbf{Level 2 / 3 / 4 / 5} &
360 / 0 / 4 / 12 \\
\midrule
\multicolumn{2}{c}{\textbf{fastexcel}} \\
\midrule
\textbf{Total commits} & 500 \\
\textbf{Commit range} &
\href{https://github.com/dhatim/fastexcel/commit/b9785779ee140ab295a03188f95ef4b6a0236a9e}{b978577}
--- 
\href{https://github.com/dhatim/fastexcel/commit/66a9dcb6022286d99f6d7252e161e4e58baa7bd9}{66a9dcb} \\
\textbf{Commit dates} & (2020-04-24) --- (2025-05-18) \\
\textbf{Significant changes} & 113 (61 regressions) \\
\midrule
\textbf{Mean energy} & 812.75 J \\
\textbf{Median energy} & 801.75 J \\
\textbf{Std. dev.} & 198.54 J \\
\midrule
\textbf{Normal dist.} & 439 / 500 (87.8\%) \\
\textbf{Outliers removed} & 38 \\
\textbf{Relative change $\Delta\%$} &
104 minor, 3 moderate, 6 major \\
\textbf{Practical significance $\Delta J$} &
108 info, 1 warning, 4 critical \\
\textbf{Level 2 / 3 / 4 / 5} &
98 / 8 / 5 / 2 \\
\bottomrule
\end{tabular}
    \end{scriptsize}
\end{table}

\begin{figure}[t!]
    \centering
    \subfigure[Cusum plot for \texttt{univocity-parsers}\label{fig:cusum-plot-univocity-parsers}]{\includegraphics[width=0.84\textwidth]{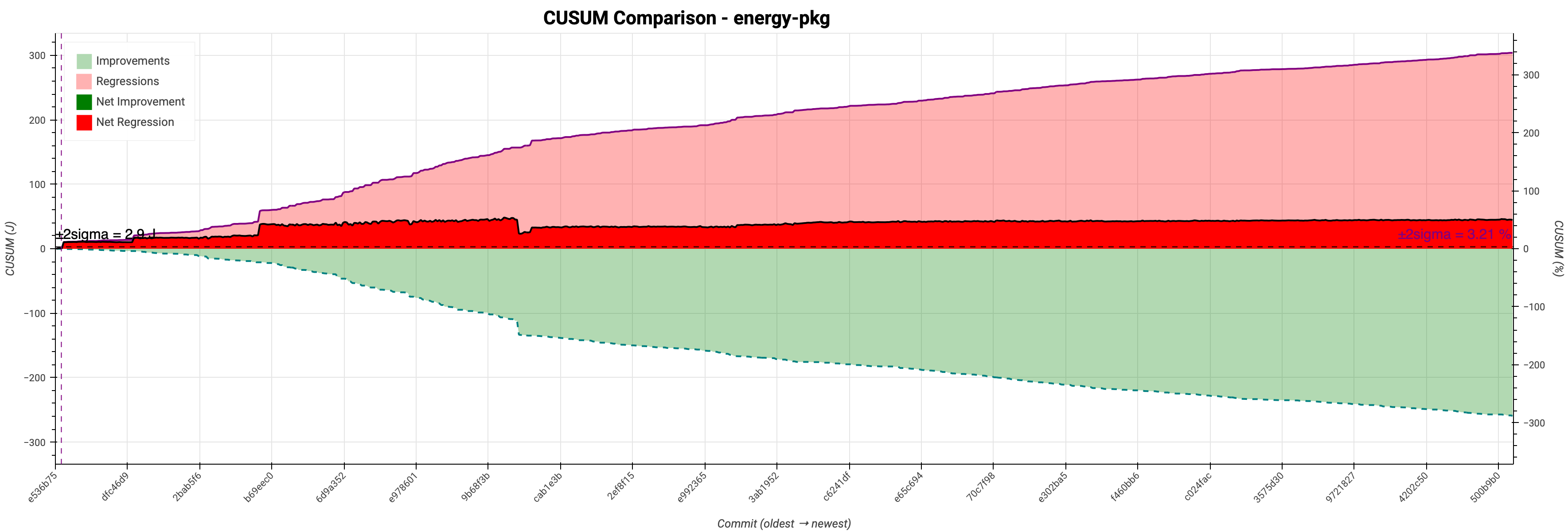}}
    \subfigure[Change points detection plot for \texttt{univocity-parsers}\label{fig:change-points-univocity-parsers}]{\includegraphics[width=0.84\textwidth]{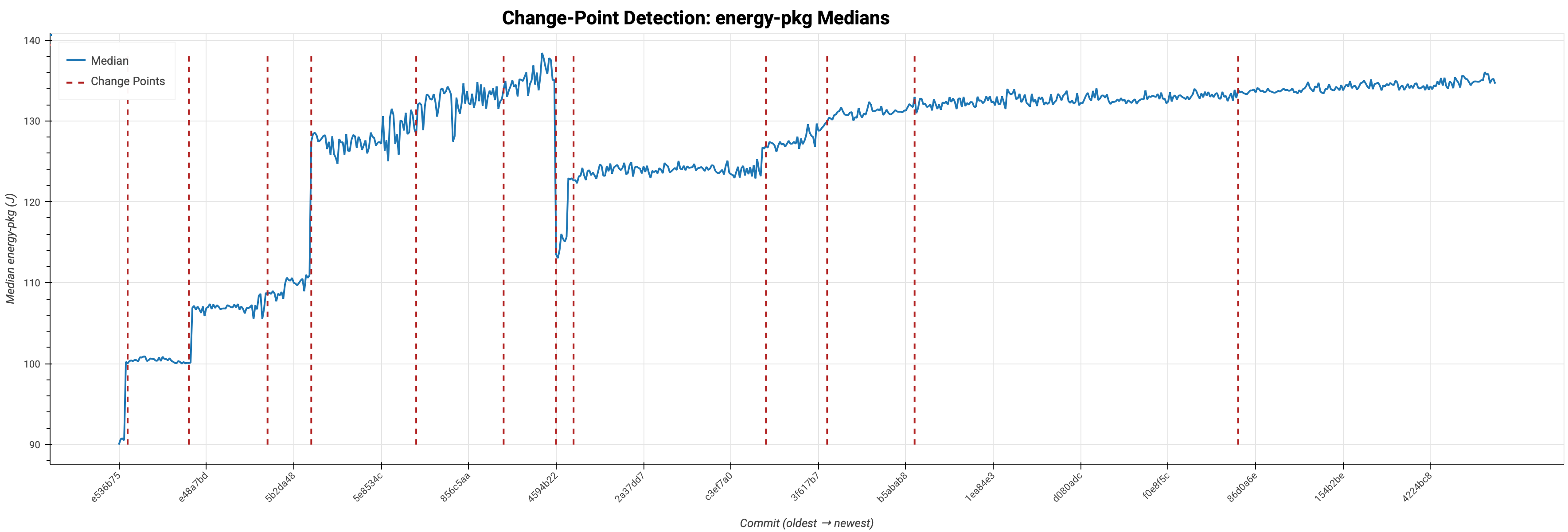}}
    \subfigure[Cusum plot for \texttt{fastexcel}\label{fig:cusum-plot-fastexcel}]{\includegraphics[width=0.84\textwidth]{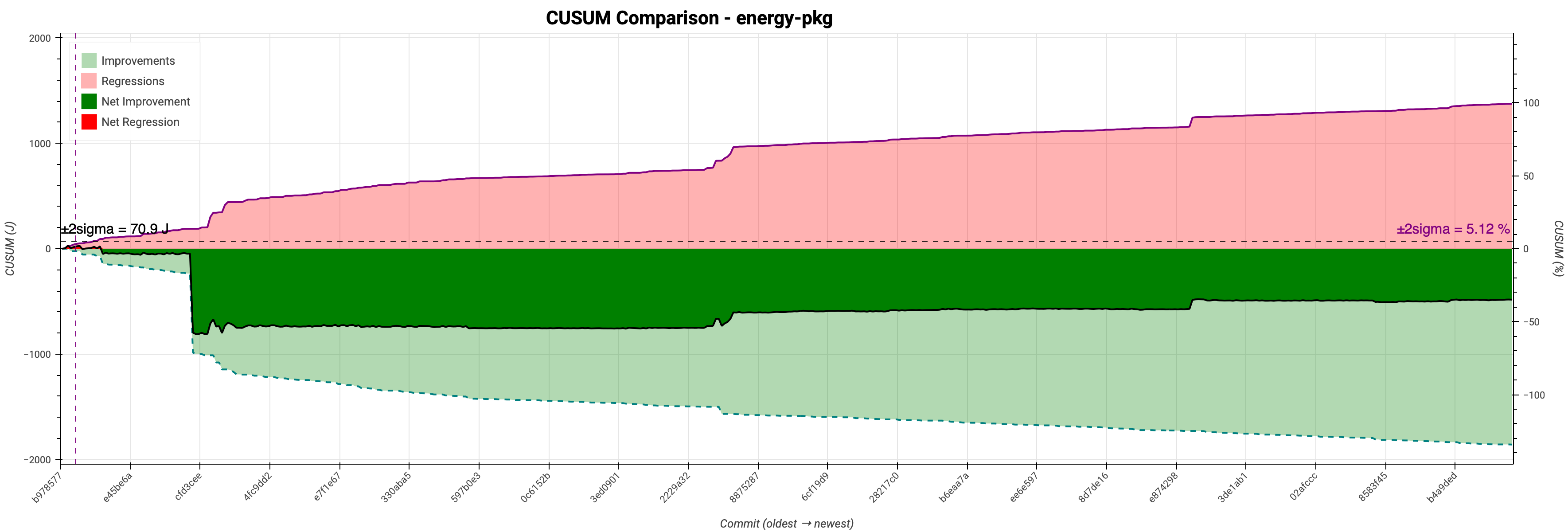}}
    \subfigure[Change points detection plot for \texttt{fastexcel}\label{fig:change-points-fastexcel}]{\includegraphics[width=0.84\textwidth]{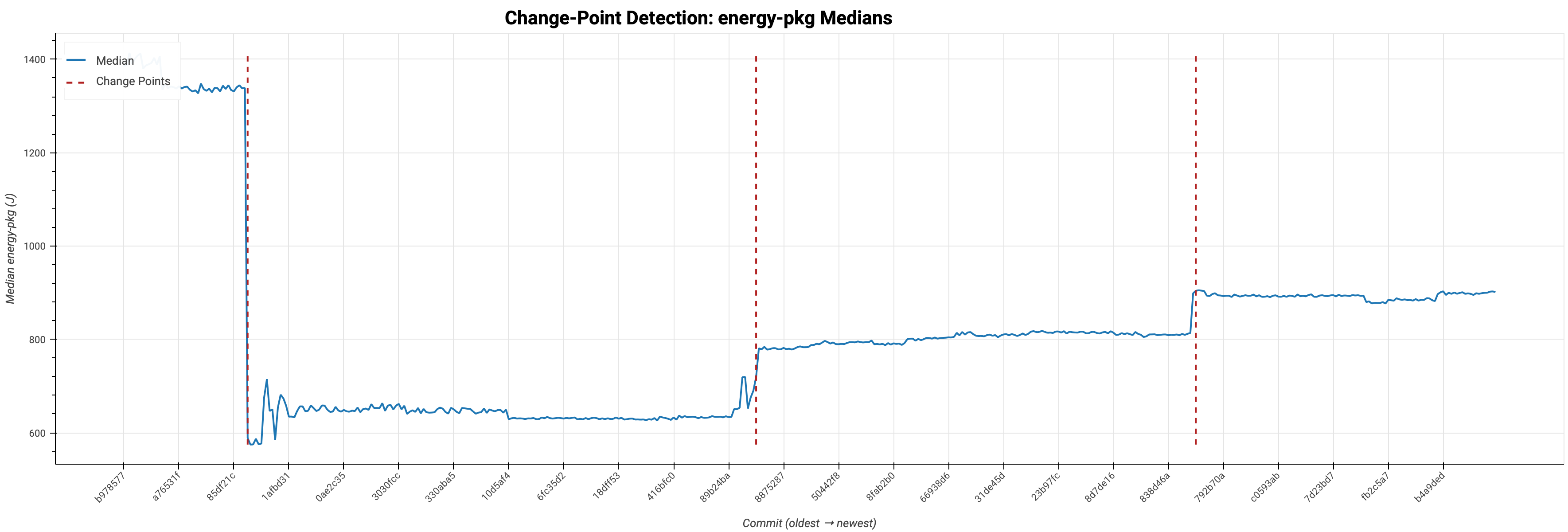}}
    \caption{Cusum and change point detection plots for the \texttt{univocity-parsers} \texttt{fastexcel} projects.}
    \label{fig:plotsresults}
\end{figure}

Table \ref{tab:case-studies-summary} contains the results summary for each project. The \texttt{fastexcel} project, although the smallest (cf. Table \ref{tab:project_overview}), consumes the most energy (812.75 joules on average). It is also the one with the lowest number of level 5 changes. One is an improvement, as confirmed by the massive drop that can be observed in figure \ref{fig:change-points-fastexcel}: commit \href{https://github.com/dhatim/fastexcel/commit/0cfb26411e1255d2858694089c55390f772b7a2b}{ 0cfb264} (message: "\textit{e2e maven profile for running test with \textbf{memory} constraints}") decreases energy consumption from 1338.64J to 588.52J, and was marked as level 5 because of the presence of the word '\textit{memory}'. 
The other level 5 commit is commit \href{https://github.com/dhatim/fastexcel/commit/9b226e2f1b5339a8650c7357c1072e54aa74dce9}{9b226e2} (message: "\textit{[...] update poi read streaming \textbf{benchmark}}"), increasing energy from 644.05J to 651.97J. This small increase is confirmed by the \textit{minor} relative change ($\Delta\%$) and practical significance $\Delta{}J$ set to \textit{info}. However, the commit message contains the word '\textit{benchmark}', one of the default words leading to labeling a commit as level 5. It illustrates the strategy,  described in the change detection of Section \ref{sec:regression-detection-phase}, specifying that levels are not necessarily cumulative. In this case, a change in the software's benchmarking policy impacting energy consumption could lead to energy issues, hence level 5, requiring attention from the developers.

Figures \ref{fig:plotsjsoup} and \ref{fig:plotsresults} present the measurements for the three project histories using the cumulated sum (cusum plot) and change point detection. We can see that overall, energy consumption tends to increase throughout the project's history. Some commits also drastically reduce the energy consumption in the project. For instance, in Figure \ref{fig:change-points-univocity-parsers}, commit \href{https://github.com/uniVocity/univocity-parsers/commit/4594b22b0bd4cb716954e946b48e88dc66418b2e}{4594b22} (message: "\textit{Reducing memory consumption [...]}") has a significant impact on the energy consumption, decreasing from 135.07J on average to 113.51J. 

\begin{tcolorbox}[enhanced, sharp corners=south,
        colback=gray!5, colframe=gray!30, boxrule=0pt,
        borderline west={2pt}{0pt}{gray},
        left=6pt, right=2pt, top=2pt, bottom=2pt]
    \textbf{Answer to RQ1}: Of the 1,944 commits analyzed for the \texttt{Jsoup} project, 92.64\% of them passed the Shapiro-Wilk normality test: 216 significant changes were detected, and 124 were classified as regressions. 
    From the 788 commits of \texttt{univocity-parser}, 87.69\% passed the Shapiro-Wilk normality test: 381 were significant, and 189 were classified as regression.
    For \texttt{fastexcel}, out of the 500 commits analyzed, 87.8\% passed the Shapiro-Wilk normality test: 113 were significant, and 61 were classified as regression.
    In summary, the \EnergyTrackr approach allows for producing stable and consistent energy values across repeated test executions, as demonstrated by a high rate of normal distributions. The data and report can serve as a strong basis for energy regression identification.
\end{tcolorbox}

\subsection{Change patterns (RQ2)}
\label{sec:case-study-insights}

To answer RQ2, we selected 20 commits flagged as significant in Table \ref{tab:case-studies-summary} that are not related to changes in the tests and manually inspected them. Significant commits stemming from test changes are interesting because, as we measure the energy consumption of the test suite, it is a good indicator that the change detection logic is effective. However, such commits are not interesting in our case as they can be trivially explained. 
We also ignored significant changes that had a low impact: i.e., that are not practically significant (\textit{info}) or have a low percentage change (\textit{minor}).

\begin{table}[t]
    \caption{Identified recurring change patterns}
    \label{tab:identified-patterns}
    \centering
    \begin{scriptsize}
        \begin{tabularx}{\linewidth}{@{}p{25mm}XX@{}}
\toprule
\textbf{Pattern}
    & \textbf{Description}
    & \textbf{Occurrences}
    \\
\midrule

\textbf{P1} --- Eager allocation   
    & Pre-allocating large buffers or object pools instead of allocating on demand.
    & \texttt{univocity-parsers} (\href{https://github.com/uniVocity/univocity-parsers/commit/4594b22b0bd4cb716954e946b48e88dc66418b2e}{4594b22})
    \\
\textbf{P2} --- Missing early exits                         
    & Absence of guard clauses or base cases in loops or methods.
    & \texttt{Jsoup} (\href{https://github.com/jhy/jsoup/commit/d0df41957bb9f1530c9cb6678e2b59fa8926ae81}{d0df419}),
    \texttt{u\-ni\-vo\-ci\-ty-par\-sers} (\href{https://github.com/uniVocity/univocity-parsers/commit/d67b605c3a84cb0d56631d0c4d0cccf2dc39db54}{d67b605}, \href{https://github.com/uniVocity/univocity-parsers/commit/403ca2edf999e67944c36b55263fc5eb2212eacc}{403ca2e})
    \\ 
\textbf{P3} --- Redundant recomputation
    & Recomputing large structures or normalizations instead of delta-updating.
    & \texttt{Jsoup} (\href{https://github.com/jhy/jsoup/commit/d80275e16ebd34bae5b48f29f3e4437e1b207955}{d80275e})
    \\ 
\textbf{P4} --- Double map lookup                    
    & Usage of \texttt{containsKey} followed by \texttt{put}, instead of a single atomic check-update.
    & \texttt{Jsoup} (\href{https://github.com/jhy/jsoup/commit/73e23c1aafe283c227b22993a9c9510d1113ac86}{73e23c1})
    \\
\textbf{P5} --- Missed reuse opportunities                    
    & Re-instantiating data structures like \texttt{StringBuilder} instead of reusing existing ones.
    & \texttt{Jsoup} (\href{https://github.com/jhy/jsoup/commit/9d9e53c8f5ab7d5dfa689ff39ee9e5dc5f7a881e}{9d9e53c})
    \\
\textbf{P6} --- Dependency upgrades                    
    & Modifications to external libraries or plugins that significantly shift energy profiles.
    &\texttt{Jsoup} (\href{https://github.com/jhy/jsoup/commit/89de79670c77039a99177fe6f739ea6f675a3a59}{89de796}, \href{https://github.com/jhy/jsoup/commit/300c521fcc3fce49a199a59f672a62111ee27171}{300c521}, \href{https://github.com/jhy/jsoup/commit/e67e5fec16200856d437c98506f19b832c8f6a70}{e67e5fe}, \href{https://github.com/jhy/jsoup/commit/2c8b57892849485ced7b946221b1eb3194e03529}{2c8b578}, \href{https://github.com/jhy/jsoup/commit/32064f089cdd59e3ed3f54bff3e160d94523d5c7}{32064f0}, \href{https://github.com/jhy/jsoup/commit/5af8413f0228332f79344b7e4233bc605b232b44}{5af8413})
    \\
\bottomrule
\end{tabularx}
    \end{scriptsize}
\end{table}

Of these 20 significant commits, 13 exhibited recurring structural or algorithmic patterns likely contributing to unnecessary energy consumption. These patterns are described in Table \ref{tab:identified-patterns}. An example of commit analysis for each identified pattern is described below. The manual analysis was done by the first author and cross-checked by the second author. Complete statistics with the related commits for each project can be found in the replication package~\cite{energytrackrdata}.

\paragraph{\textbf{P1} --- Eager allocation}

Commit \texttt{4594b22} from \texttt{univocity-parsers} is classified as Level 5 and offers a 16\% savings in energy consumption. The commit diff with the old code (in red) and new code (in green) is:

\noindent
\begin{minipage}[t]{0.49\textwidth}
\vspace{0pt}
\begin{deletedcode}
\begin{lstlisting}
for (int i = 0; i < size; i++) {
	instancePool[i] = new Entry<T>(newInstance(), i);
	instanceIndexes[i] = i;
}
\end{lstlisting}
\end{deletedcode}
\end{minipage}
\hfill
\begin{minipage}[t]{0.49\textwidth}
\vspace{0pt}
\begin{addedcode}
\begin{lstlisting}
Arrays.fill(instanceIndexes, -1);
instancePool[0] = new Entry<T>(newInstance(), 0);
instanceIndexes[0] = 0;
\end{lstlisting}
\end{addedcode}
\end{minipage}
\smallskip

The pattern \texttt{Eager allocation} can be deduced from this code, and saves energy because it avoids excessive memory usage and reduces CPU usage spent on unused object creation.

\paragraph{\textbf{P2} --- Missing early exits}

Commit \texttt{d67b605}, from \texttt{univocity-parsers}, is classified as Level 3, and offers a 4.3\% savings in energy consumption. From the commit diff, we can retrieve the following code:

\noindent
\begin{minipage}[t]{0.49\textwidth}
\vspace{0pt}
\begin{deletedcode}
\begin{lstlisting}
protected final int skipLeadingWhitespace(String element) {
    for (int i = 0; i < element.length(); i++) {
        char nextChar = element.charAt(i);
        if (!(nextChar <= ' ')) {
            return i;
        }
    }
    return 0;
}
\end{lstlisting}
\end{deletedcode}
\end{minipage}
\hfill
\begin{minipage}[t]{0.49\textwidth}
\vspace{0pt}
\begin{addedcode}
\begin{lstlisting}
protected final int skipLeadingWhitespace(String element) {
    if(element.isEmpty()){
        return 0;
    } 
    for (int i = 0; i < element.length(); i++) {
        char nextChar = element.charAt(i);
        if (!(nextChar <= ' ')) {
            return i;
        }
    }
    return element.length() - 1;
}
\end{lstlisting}
\end{addedcode}
\end{minipage}
\smallskip

The old code did not check if strings only contained white space characters, so the old code was processing them, which was useless instead of returning. The new code returns early, which saves energy.

\paragraph{\textbf{P3} --- Redundant recomputation}
Commit \texttt{d80275e}, from \texttt{Jsoup}, is classified as Level 2, and offers a 3\% savings in energy consumption.

\noindent
\begin{minipage}[t]{0.49\textwidth}
\vspace{0pt}
\begin{deletedcode}
\begin{lstlisting}
append = append.replace(TokeniserState.nullChar, Tokeniser.replacementChar);
tagName = tagName == null ? append : tagName.concat(append);
normalName = ParseSettings.normalName(tagName);
\end{lstlisting}
\end{deletedcode}
\end{minipage}
\hfill
\begin{minipage}[t]{0.49\textwidth}
\vspace{0pt}
\begin{addedcode}
\begin{lstlisting}
append = append.replace(TokeniserState.nullChar, Tokeniser.replacementChar);
tagName = tagName == null ? append : tagName.concat(append);
// perf: normalize just the appended content
String normalAppend = ParseSettings.normalName(append);
normalName = normalName == null ? normalAppend : normalName.concat(normalAppend);
\end{lstlisting}
\end{addedcode}
\end{minipage}
\smallskip

The old code was unnecessarily normalizing the full \texttt{tagName} instead of focusing on the relevant sub-part (\texttt{normalAppend}). Depending on the context, applying treatment multiple times to the same data could be the correct behavior. However, raising the developers' attention to numerous (same) processing on the same data could lead to energy consumption reduction.

\paragraph{\textbf{P4} --- Double map lookup}
Commit \texttt{73e23c1}, from \texttt{Jsoup}, is classified as Level 2 and offers a 2.7\% savings in energy consumption. 

\noindent
\begin{minipage}[t]{0.49\textwidth}
\vspace{0pt}
\begin{deletedcode}
\begin{lstlisting}
for (Element root : roots) {
    final Elements found = select(evaluator, root);
    for (Element el : found) {
        if (!seenElements.containsKey(el)) {
            elements.add(el);
            seenElements.put(el, Boolean.TRUE);
        }
    }
}
\end{lstlisting}
\end{deletedcode}
\end{minipage}
\hfill
\begin{minipage}[t]{0.49\textwidth}
\vspace{0pt}
\begin{addedcode}
\begin{lstlisting}
for (Element root : roots) {
    final Elements found = select(evaluator, root);
    for (Element el : found) {
        if (seenElements.put(el, Boolean.TRUE) == null) {
            elements.add(el);
        }
    }
}
\end{lstlisting}
\end{addedcode}
\end{minipage}
\smallskip

The developer appears to be aware of potential performance improvement, as indicated by the commit message "\textit{Minor performance improvement to Selector.select()}". In addition to energy improvements, this pattern seems to enhance the overall performance. The primary modification involves performing the lookup (call to the \texttt{put} method) only once within the conditional statement, thereby reducing CPU overhead. Merging the body statement with the condition statement can reduce redundant computation.


\paragraph{\textbf{P5} --- Missed reuse opportunities}

Commit \texttt{9d9e53c}, from \texttt{Jsoup}, is classified as Level 2 and offers a 3\% savings in energy consumption. The diff shows that the code was creating useless string builders. The new code reuses the already existing builder instead. 

\noindent
\begin{minipage}[t]{0.49\textwidth}
\vspace{0pt}
\begin{deletedcode}
\begin{lstlisting}
if (charsBuilder.length() > 0) {
    String str = charsBuilder.toString();
    charsBuilder.delete(0, charsBuilder.length());
}
\end{lstlisting}
\end{deletedcode}
\end{minipage}
\hfill
\begin{minipage}[t]{0.49\textwidth}
\vspace{0pt}
\begin{addedcode}
\begin{lstlisting}
final StringBuilder cb = this.charsBuilder;
if (cb.length() != 0) {
    String str = cb.toString();
    cb.delete(0, cb.length());
}
\end{lstlisting}
\end{addedcode}
\end{minipage}
\smallskip

\paragraph{\textbf{P6} --- Dependency upgrades}

Commit \texttt{89de796}, from \texttt{Jsoup}, is classified as Level 4. Here, changing the version of \texttt{junit-jupiter} increased the energy consumption of running the same test by 6.1\%. Similar behavior is observed for the other commits listed in Table \ref{tab:identified-patterns}.

\begin{tcolorbox}[enhanced, sharp corners=south,
        colback=gray!5, colframe=gray!30, boxrule=0pt,
        borderline west={2pt}{0pt}{gray},
        left=6pt, right=2pt, top=2pt, bottom=2pt]
    \textbf{Answer to RQ2}: The manually analyzed commits flagged as regressions exhibit recurring patterns that likely contribute to unnecessary energy consumption. From these candidate patterns, at least two stand out: patterns linked to dependency changes that sometimes cause regressions and patterns due to missed early exits in code that could save a considerable amount of energy. 
\end{tcolorbox}

\section{Threats to validity}
\label{sec:threats-to-validity}

The section presents the threats to validity related to our qualitative and quantitative evaluation of \EnergyTrackr.

\subsection{Internal validity}

\paragraph{Energy measurement}

Energy measurements are subject to many sources of variability. Multiple precautions were taken to mitigate this, including repeated measurements and randomized test execution. Despite those efforts, some measurement noise may have occurred during measurements or hardware variability, like a room temperature increase, which was not controlled during this experimentation. However, thermal control was done inside the measurement pipeline to avoid thermal throttling via the temperature check stage. Furthermore, utilizing RAPL for energy measurements presents limitations, as it only monitors the entire CPU. Consequently, this approach does not yield finer-grained data that could facilitate the identification of specific sources of performance regressions, such as memory, storage devices, or network adapters, that may contribute to increased energy consumption.

Another factor that may have impacted results is the choice of the batch size. On the one hand, smaller batches may reduce the benefit of randomization. On the other hand, bigger batches increase the time between the first measurement for a commit and its last measurement, which may increase the variance caused by environmental noise. This effect was not explicitly measured in this study, but it remains a potential noise source worth investigating in future experiments.

An improvement that could be made to \EnergyTrackr would be to assess the normality of measurements directly within the measurement pipeline, and automatically schedule non-normal commits for re-measurement. This would improve the results and ensure a higher number of normally distributed measurements in the final report.

Regarding \EnergyTrackr's implementation, we took the following precautions: it comes with unit tests (the project aims for 80\% to 100\% line coverage of test coverage for the measurement modules). We also used linters (Ruff, Pylint, and Pyright) to identify potential defects. Finally, we followed software engineering best practices with GitHub Actions, checking that all the formatting and linting rules are respected and verifying that the code coverage passes the defined threshold and that the documentation builds.

\paragraph{Changes in the test suite}

Following the idea of \citet{Danglot2024CanWeSpot}, we assume that running tests can detect energy regressions, especially when good test coverage is in place. 
While test-based measurements have the downside of pointing to biased results that are (just) test modifications and that must be ignored, they have the advantage of eliminating the need for manual creation of custom tests, thereby facilitating scalability without requiring extensive project-specific expertise.
However, the number of passing tests was not measured. This could imply that some commits spotted as regressions could be because some tests fail (which is also a regression, but not one related to energy consumption). For our manual analysis, we filtered out commits related to changes in the test suite to avoid such effects.
Another approach for our quantitative evaluation would have been to rely on mutation testing \cite{papadakis_mutation_2019} and artificially introduce known regressions within the test-covered codebase to measure precision and recall accurately. Instead, we rely on open-source projects to increase the impact and relevance of our evaluation.

\subsection{External validity}

The projects evaluated in this study varied significantly in their application domains. However, the limited number of projects examined may affect the representativeness of the findings. 
We selected Git-based open-source projects from the \textit{Awesome-Java} list to enable reproducibility while ensuring relevance based on the project's popularity. We considered only projects with a certain commit history of at least 500 commits. While this selection relies on practical considerations for our research, a larger and more diverse set of benchmark projects would be required to achieve broader generalizability. The chosen projects are characterized by dependency management and good test suites, which may not reflect the typical characteristics observed in the average Git repository.

\subsection{Construct validity}

The tool primarily emphasizes the short-term execution of the test suite, which may not fully reflect real-world software usage scenarios. Nonetheless, we argue that assessing the energy consumption of the test suite provides valuable insights that can inform efforts that must be performed to enhance the sustainability of software systems.

\section{Discussion}
\label{sec:discussion}


\subsection{Energy consumption-aware linter}
\label{sec:findings-interpretation}

Several static analysis tools exist, like linters, that developers can use to spot potential issues related to energy consumption. Such tools, however, usually require data about regression defects to ensure their relevance to the state of the practice. The patterns identified in our manual analysis (RQ2) can serve as a basis for building such a linter and further ease the triage of the different commits flagged as regressions. 
For instance, from Table \ref{tab:identified-patterns} analysis, P2 and P6 appear to be recurring patterns in the three projects. Although more data are needed to support (and generalize) them, we can derive a first set of rules based on the patterns of Table \ref{tab:identified-patterns}. Extending our analysis to devise such rules is part of our future work.
Based on the six observed patterns described in Table \ref{tab:identified-patterns}, we outline the following (proto-)rules: 
\begin{compactitem}
    \item[\textbf{R1 (P1)}:] Warn when allocating full object pools at once and suggest lazy initialization instead.
    \item[\textbf{R2 (P2)}:] Flag loops or functions lacking cheap guard conditions.
    \item[\textbf{R3 (P3)}:] Warn when entire structures are recomputed instead of incrementally updated.
    \item[\textbf{R4 (P4)}:] Detect \texttt{containsKey} or \texttt{put} patterns and recommend replacement.
    \item[\textbf{R5 (P5)}:] Warn on re-instantiation of temporary buffers where reuse is possible.
\end{compactitem}
\textbf{P6} cannot be turned into a rule prototype for a linter, as code refactoring techniques cannot improve dependency upgrades. However, raising developers' awareness about the potential impact of the upgrade could be considered with finer-grained dependency call analysis like the one from \citet{hejderup_prazi_2022}.

An interesting finding is that for some commits, e.g., commit \href{https://github.com/uniVocity/univocity-parsers/commit/4594b22b0bd4cb716954e946b48e88dc66418b2e}{4594b22} for \texttt{univocity-parsers}, developers were aware of the potential performance impact, but for other commits, e.g., for commit \href{https://github.com/jhy/jsoup/commit/89de79670c77039a99177fe6f739ea6f675a3a59}{89de796} for \texttt{Jsoup}, they were not. This highlights the need for more awareness among the developer community, which energy-aware linters would help address without requiring longer dynamic analysis. 
Another observation is that, based on the commit messages, dependency updates frequently lead to changes in energy consumption, often without the developers being fully aware of the impact on their test suite or the overall software performance.

\subsection{Practical usages for developers}
\label{sec:implications-developers}

With \EnergyTrackr, developers can monitor energy regressions and improvements at the commit, branch, or tag level with high precision. The implementation offers a user-friendly report interface and a fully configurable system that supports various projects. Users can measure the energy consumption of their test suite, specific test cases designed to evaluate application behavior, or any command they choose. This analysis can be a one-shot or repeated regularly to track the project's energy consumption evolution.

While the identified anti-patterns provide a valuable starting point for future development of energy-aware linters, further research is necessary to fully realize the practical benefits for developers. Nonetheless, these anti-patterns can be shared within the community to help highlight recurring bad practices.
\EnergyTrackr is designed for seamless integration into existing workflows with minimal overhead. While detecting energy regressions can be automated through scripting, our current implementation is not yet integrated into CI workflows, e.g., GitHub Actions.

\subsection{Limitations of the current implementation}
\label{sec:limitations}

\paragraph{Measurement limitations}

Using \texttt{perf} introduces some limitations, including dependence on RAPL support, which may not be available on all hardware configurations, and the need for specific permissions that are typically restricted in CI environments. Additionally, \texttt{perf} is exclusively available on Linux platforms, limiting its applicability to Linux-based systems. A further limitation of the RAPL interface is its inability to provide granular energy consumption data for individual CPU cores. Although heuristics exist to estimate per-core energy usage, these methods were not employed in this analysis, as they may reduce overall measurement accuracy compared to the raw RAPL data.

\paragraph{Energy regression detection}

The current algorithm is straightforward: it sequentially traverses the entire commit history. While this approach effectively identifies all potential energy fluctuations, it is relatively time-intensive and may include examining extensive commit ranges with minimal or no significant energy variation. In our future work, we plan to devise a more efficient algorithm, based on the dichotomic search, that will take measurements into account to slice the commit history and reduce the number of commits to analyze.
\section{Conclusion and Future Work}
\label{sec:conclusion}

In this paper, we designed, implemented, and evaluated an energy measurement and reporting tool: \EnergyTrackr. Its modular implementation leverages \texttt{perf}, which relies on RAPL performance counters to accurately detect regressions in the commit history of projects. Using statistical metrics and their corresponding categories, our approach successfully identified significant regressions in multiple commits. After the manual analysis of commits flagged as energy regressions by \EnergyTrackr, multiple recurring code change patterns, such as missing early exits or dependency upgrades, were linked to significant energy change.

\EnergyTrackr provides a comprehensive view for monitoring energy regressions and improvements at the commit, branch, or tag level with high accuracy. It features an intuitive reporting interface and a highly configurable system for various projects. Our manual analysis showed that a significant portion of energy regressions stems from dependency upgrades, highlighting the need for developers to monitor their libraries' impacts across diverse use cases. Key energy anti-patterns include missing early-exit conditions and unnecessary computations. However, more extensive research is needed to build a full-featured energy-aware linter with more patterns and to confirm that the discovered patterns are indeed present in many projects. 
    


There are multiple tracks for improvement in this work. \EnergyTrackr could be extended to other programming languages, as it is language-agnostic. This extension could determine whether specific anti-patterns are consistent across languages. We could also rely on automated pattern mining to find anti-patterns from regression commits, easing the analysis for researchers and developers. Such patterns could then serve as input to devise rules for energy-aware linters and provide developers with automated warnings about potential energy regressions. 

\EnergyTrackr's implementation also has some limitations. Currently, its granularity is limited to the commit level, which requires users to manually inspect whole commits to identify the cause of energy regressions. Since \texttt{perf} allows for high-frequency sampling, enhancing the tool to operate at the function level is theoretically possible by implementing, for instance, JoularJX's approach. The energy regression detection strategy also requires building and executing large portions of the commit history. Our future work includes designing a new algorithm based on the dichotomic search to reduce the number of commits to analyze drastically. 

The different visualizations implemented for \EnergyTrackr helped us identify code patterns during our manual analysis. Extending that analysis will help design a linter to detect energy regressions from the code statically. Finally, the following steps include a full-scale evaluation of \EnergyTrackr with developers to evaluate the practical usability of our implementation for energy regression identification and, more importantly, debugging.

\section*{Data Availability}

Our data are openly available in our replication package \cite{energytrackrdata}. Our implementation of \EnergyTrackr is available as an open-source project \cite{energytrackr}.

\section*{Acknowledgments}

This research was partially funded by the CyberExcellence by DigitalWallonia project (No. 2110186) funded by the Public Service of Wallonia (SPW Recherche).

\bibliographystyle{plainnat}
\bibliography{references}

\begin{thebibliography}{54}
\providecommand{\natexlab}[1]{#1}
\providecommand{\url}[1]{\texttt{#1}}
\expandafter\ifx\csname urlstyle\endcsname\relax
  \providecommand{\doi}[1]{doi: #1}\else
  \providecommand{\doi}{doi: \begingroup \urlstyle{rm}\Url}\fi

\bibitem[Acar et~al.(2016)Acar, Alptekin, Gelas, and Ghodous]{Acar2016ImpactSourceCode}
Hayri Acar, G{\"u}lfem~I Alptekin, Jean-Patrick Gelas, and Parisa Ghodous.
\newblock The {{Impact}} of {{Source Code}} in {{Software}} on {{Power Consumption}}.
\newblock \emph{Int. J. Electron. Bus. Manag.}, 14:\penalty0 42--52, 2016.

\bibitem[Arcuri and Briand(2014)]{Arcuri2014HitchhikersGuideStatistical}
Andrea Arcuri and Lionel Briand.
\newblock A {{Hitchhiker}}'s guide to statistical tests for assessing randomized algorithms in software engineering.
\newblock \emph{Software Testing, Verification and Reliability}, 24\penalty0 (3):\penalty0 219--250, 2014.
\newblock \doi{10.1002/stvr.1486}.

\bibitem[Banerjee et~al.(2014)Banerjee, Chong, Chattopadhyay, and Roychoudhury]{Banerjee2014DetectingEnergyBugs}
Abhijeet Banerjee, Lee~Kee Chong, Sudipta Chattopadhyay, and Abhik Roychoudhury.
\newblock Detecting energy bugs and hotspots in mobile apps.
\newblock In \emph{Proceedings of the 22nd {{ACM SIGSOFT International Symposium}} on {{Foundations}} of {{Software Engineering}}}, pages 588--598, Hong Kong China, November 2014. ACM.
\newblock ISBN 978-1-4503-3056-5.
\newblock \doi{10.1145/2635868.2635871}.

\bibitem[Bangash et~al.(2017)Bangash, Sahar, and Beg]{Bangash2017MethodologyRelatingSoftware}
Abdul~Ali Bangash, Hareem Sahar, and Mirza~Omer Beg.
\newblock A {{Methodology}} for {{Relating Software Structure}} with {{Energy Consumption}}.
\newblock In \emph{17th {{IEEE International Working Conference}} on {{Source Code Analysis}} and {{Manipulation}}, {{SCAM}} 2017, {{Shanghai}}, {{China}}, {{September}} 17-18, 2017}, pages 111--120. IEEE Computer Society, September 2017.
\newblock \doi{10.1109/SCAM.2017.18}.

\bibitem[Bechet(2025{\natexlab{a}})]{energytrackr}
{Fran\c{c}ois} Bechet.
\newblock Energytrackr, 2025{\natexlab{a}}.
\newblock URL \url{https://github.com/snail-unamur/energytrackr}.

\bibitem[Bechet(2025{\natexlab{b}})]{energytrackrdata}
{Fran\c{c}ois} Bechet.
\newblock Replication package, 2025{\natexlab{b}}.
\newblock URL \url{https://doi.org/10.5281/zenodo.15594507}.

\bibitem[Bonvoisin et~al.(2024)Bonvoisin, Quinton, and Rouvoy]{Bonvoisin2024UnderstandingPerformanceEnergyTradeoffs}
Alexandre Bonvoisin, Cl{\'e}ment Quinton, and Romain Rouvoy.
\newblock Understanding the {{Performance-Energy Tradeoffs}} of {{Object-Relational Mapping Frameworks}}.
\newblock In \emph{{{SANER}}'24 - 31th {{IEEE International Conference}} on {{Software Analysis}}, {{Evolution}} and {{Reengineering}}}, page~11. IEEE, March 2024.

\bibitem[Cohen(1988)]{Cohen1988}
Jacob Cohen.
\newblock \emph{Statistical power analysis for the behavioral sciences}.
\newblock routledge, 1988.
\newblock \doi{10.4324/9780203771587}.

\bibitem[Connolly~Bree and {\'O}~Cinn{\'e}ide(2025)]{ConnollyBree2025HowSoftwareDesign}
D{\'e}agl{\'a}n Connolly~Bree and Mel {\'O}~Cinn{\'e}ide.
\newblock How {{Software Design Affects Energy Performance}}: {{A Systematic Literature Review}}.
\newblock \emph{Journal of Software: Evolution and Process}, 37\penalty0 (4):\penalty0 e70014, 2025.
\newblock ISSN 2047-7481.
\newblock \doi{10.1002/smr.70014}.

\bibitem[Cruz(2021)]{Cruz2021GreenSoftwareEngineering}
Lu{\'i}s Cruz.
\newblock Green {{Software Engineering Done Right}}: A {{Scientific Guide}} to {{Set Up Energy Efficiency Experiments}}.
\newblock https://luiscruz.github.io/2021/10/10/scientific-guide.html, 2021.

\bibitem[Danglot et~al.(2024)Danglot, Falleri, and Rouvoy]{Danglot2024CanWeSpot}
Benjamin Danglot, Jean-R{\'e}my Falleri, and Romain Rouvoy.
\newblock Can we spot energy regressions using developers tests?
\newblock \emph{Empirical Software Engineering}, 29\penalty0 (5):\penalty0 121, July 2024.
\newblock ISSN 1573-7616.
\newblock \doi{10.1007/s10664-023-10429-1}.

\bibitem[Dutoit(2012)]{Dutoit2012GraphicalExploratoryData}
S.~H.~C. Dutoit.
\newblock \emph{Graphical Exploratory Data Analysis}.
\newblock Springer, 2012.
\newblock ISBN 1-4612-9371-5 978-1-4612-9371-2.

\bibitem[Efron and Tibshirani(1993)]{Efron1993IntroductionBootstrap}
Bradley Efron and Robert Tibshirani.
\newblock \emph{An Introduction to the Bootstrap}.
\newblock Springer, 1993.
\newblock ISBN 978-1-4899-4541-9.
\newblock \doi{10.1007/978-1-4899-4541-9}.

\bibitem[Fahad et~al.(2019)Fahad, Shahid, Manumachu, and Lastovetsky]{Fahad2019ComparativeStudyMethods}
Muhammad Fahad, Arsalan Shahid, Ravi~Reddy Manumachu, and Alexey Lastovetsky.
\newblock A {{Comparative Study}} of {{Methods}} for {{Measurement}} of {{Energy}} of {{Computing}}.
\newblock \emph{Energies}, 12\penalty0 (11):\penalty0 2204, January 2019.
\newblock ISSN 1996-1073.
\newblock \doi{10.3390/en12112204}.

\bibitem[Fieni et~al.(2024)Fieni, Acero, Rust, and Rouvoy]{Fieni2024PowerAPIPythonFramework}
Guillaume Fieni, Daniel~Romero Acero, Pierre Rust, and Romain Rouvoy.
\newblock {{PowerAPI}}: {{A Python}} framework for building software-defined power meters.
\newblock \emph{Journal of Open Source Software}, 9\penalty0 (98):\penalty0 6670, June 2024.
\newblock \doi{10.21105/joss.06670}.

\bibitem[Freitag et~al.(2021)Freitag, {Berners-Lee}, Widdicks, Knowles, Blair, and Friday]{Freitag2021RealClimateTransformative}
Charlotte Freitag, Mike {Berners-Lee}, Kelly Widdicks, Bran Knowles, Gordon~S. Blair, and Adrian Friday.
\newblock The real climate and transformative impact of {{ICT}}: {{A}} critique of estimates, trends, and regulations.
\newblock \emph{Patterns}, 2\penalty0 (9):\penalty0 100340, September 2021.
\newblock ISSN 2666-3899.
\newblock \doi{10.1016/j.patter.2021.100340}.

\bibitem[Georgiou et~al.(2022)Georgiou, Kechagia, Sharma, Sarro, and Zou]{Georgiou2022GreenAIDeep}
Stefanos Georgiou, Maria Kechagia, Tushar Sharma, Federica Sarro, and Ying Zou.
\newblock Green {{AI}}: Do deep learning frameworks have different costs?
\newblock In \emph{44th {{IEEE}}/{{ACM}} 44th {{International Conference}} on {{Software Engineering}}, {{ICSE}} 2022, {{Pittsburgh}}, {{PA}}, {{USA}}, {{May}} 25-27, 2022}, {{ICSE}} '22, pages 1082--1094, New York, NY, USA, July 2022. ACM.
\newblock \doi{10.1145/3510003.3510221}.

\bibitem[H{\"a}hnel et~al.(2012)H{\"a}hnel, D{\"o}bel, V{\"o}lp, and H{\"a}rtig]{Hahnel2012MeasuringEnergyConsumption}
Marcus H{\"a}hnel, Bj{\"o}rn D{\"o}bel, Marcus V{\"o}lp, and Hermann H{\"a}rtig.
\newblock Measuring energy consumption for short code paths using {{RAPL}}.
\newblock \emph{SIGMETRICS Perform. Evaluation Rev.}, 40\penalty0 (3):\penalty0 13--17, January 2012.
\newblock ISSN 0163-5999.
\newblock \doi{10.1145/2425248.2425252}.

\bibitem[Hao et~al.(2013)Hao, Li, Halfond, and Govindan]{Hao2013EstimatingMobileApplication}
Shuai Hao, Ding Li, William G.~J. Halfond, and Ramesh Govindan.
\newblock Estimating mobile application energy consumption using program analysis.
\newblock In {David Notkin}, {Betty H. C. Cheng}, and {Klaus Pohl}, editors, \emph{35th {{International Conference}} on {{Software Engineering}}, {{ICSE}} '13, {{San Francisco}}, {{CA}}, {{USA}}, {{May}} 18-26, 2013}, pages 92--101, San Francisco, CA, USA, May 2013. IEEE Computer Society.
\newblock ISBN 978-1-4673-3076-3 978-1-4673-3073-2.
\newblock \doi{10.1109/ICSE.2013.6606555}.

\bibitem[Harman et~al.(2015)Harman, Jia, and Zhang]{Harman2015AchievementsOpenProblems}
Mark Harman, Yue Jia, and Yuanyuan Zhang.
\newblock Achievements, {{Open Problems}} and {{Challenges}} for {{Search Based Software Testing}}.
\newblock In \emph{8th {{IEEE International Conference}} on {{Software Testing}}, {{Verification}} and {{Validation}}, \{\vphantom\}{{ICST}}\vphantom\{\} 2015, {{Graz}}, {{Austria}}, {{April}} 13-17, 2015}, pages 1--12. IEEE Computer Society, April 2015.
\newblock \doi{10.1109/ICST.2015.7102580}.

\bibitem[Hejderup et~al.(2022)Hejderup, Beller, Triantafyllou, and Gousios]{hejderup_prazi_2022}
Joseph Hejderup, Moritz Beller, Konstantinos Triantafyllou, and Georgios Gousios.
\newblock Präzi: from package-based to call-based dependency networks.
\newblock \emph{Empirical Software Engineering}, 27\penalty0 (5):\penalty0 102, September 2022.
\newblock \doi{10.1007/s10664-021-10071-9}.

\bibitem[Hindle et~al.(2014)Hindle, Wilson, Rasmussen, Barlow, Campbell, and Romansky]{Hindle2014GreenMinerHardwareBaseda}
Abram Hindle, Alex Wilson, Kent Rasmussen, E.~Jed Barlow, Joshua~Charles Campbell, and Stephen Romansky.
\newblock {{GreenMiner}}: A hardware based mining software repositories software energy consumption framework.
\newblock In Premkumar~T. Devanbu, Sung Kim, and Martin Pinzger, editors, \emph{11th Working Conference on Mining Software Repositories, {{MSR}} 2014, Proceedings, May 31 - June 1, 2014, Hyderabad, India}, pages 12--21. ACM, 2014.
\newblock \doi{10.1145/2597073.2597097}.

\bibitem[Hintze and Nelson(1998)]{Hintze1998ViolinPlotsBox}
Jerry~L. Hintze and Ray~D. Nelson.
\newblock Violin plots: A box plot-density trace synergism.
\newblock \emph{The American Statistician}, 52\penalty0 (2):\penalty0 181--184, 1998.
\newblock \doi{10.1080/00031305.1998.10480559}.

\bibitem[Iglewicz and Hoaglin(1993)]{Iglewicz1993Volume16How}
Boris Iglewicz and David~C. Hoaglin.
\newblock \emph{Volume 16: {{How}} to {{Detect}} and {{Handle Outliers}}}.
\newblock Quality Press, January 1993.
\newblock ISBN 978-0-87389-260-5.

\bibitem[Jay et~al.(2023)Jay, Ostapenco, Lefevre, Trystram, Orgerie, and Fichel]{Jay2023ExperimentalComparisonSoftwarebased}
Mathilde Jay, Vladimir Ostapenco, Laurent Lefevre, Denis Trystram, Anne-C{\'e}cile Orgerie, and Benjamin Fichel.
\newblock An experimental comparison of software-based power meters: Focus on {{CPU}} and {{GPU}}.
\newblock In \emph{2023 {{IEEE}}/{{ACM}} 23rd {{International Symposium}} on {{Cluster}}, {{Cloud}} and {{Internet Computing}} ({{CCGrid}})}, pages 106--118, Bangalore, India, May 2023. IEEE.
\newblock ISBN 979-8-3503-0119-9.
\newblock \doi{10.1109/CCGrid57682.2023.00020}.

\bibitem[Khan et~al.(2018)Khan, Hirki, Niemi, Nurminen, and Ou]{Khan2018RAPLActionExperiences}
Kashif~Nizam Khan, Mikael Hirki, Tapio Niemi, Jukka~K. Nurminen, and Zhonghong Ou.
\newblock {{RAPL}} in action: Experiences in using {{RAPL}} for power measurements.
\newblock \emph{ACM Trans. Model. Perform. Eval. Comput. Syst.}, 3\penalty0 (2):\penalty0 9:1--9:26, March 2018.
\newblock ISSN 2376-3639.
\newblock \doi{10.1145/3177754}.

\bibitem[Le~Goaer and Hertout(2023)]{LeGoaer2023EcoCodeSonarQubePlugin}
Olivier Le~Goaer and Julien Hertout.
\newblock {{ecoCode}}: A {{SonarQube Plugin}} to {{Remove Energy Smells}} from {{Android Projects}}.
\newblock In \emph{Proceedings of the 37th {{IEEE}}/{{ACM International Conference}} on {{Automated Software Engineering}}}, {{ASE}} '22, pages 1--4, New York, NY, USA, January 2023. Association for Computing Machinery.
\newblock ISBN 978-1-4503-9475-8.
\newblock \doi{10.1145/3551349.3559518}.

\bibitem[Li et~al.(2013)Li, Hao, Halfond, and Govindan]{Li2013CalculatingSourceLine}
Ding Li, Shuai Hao, William G.~J. Halfond, and Ramesh Govindan.
\newblock Calculating source line level energy information for {{Android}} applications.
\newblock In {Mauro Pezz{\`e}} and {Mark Harman}, editors, \emph{International {{Symposium}} on {{Software Testing}} and {{Analysis}}, {{ISSTA}} '13, {{Lugano}}, {{Switzerland}}, {{July}} 15-20, 2013}, pages 78--89, Lugano Switzerland, July 2013. ACM.
\newblock ISBN 978-1-4503-2159-4.
\newblock \doi{10.1145/2483760.2483780}.

\bibitem[Liu et~al.(2015)Liu, Pinto, and Liu]{Liu2015DataOrientedCharacterizationApplicationLevel}
Kenan Liu, Gustavo Pinto, and Yu~David Liu.
\newblock Data-{{Oriented Characterization}} of {{Application-Level Energy Optimization}}.
\newblock In Alexander Egyed and Ina Schaefer, editors, \emph{Fundamental {{Approaches}} to {{Software Engineering}}}, Lecture {{Notes}} in {{Computer Science}}, pages 316--331, Berlin, Heidelberg, 2015. Springer.
\newblock ISBN 978-3-662-46675-9.
\newblock \doi{10.1007/978-3-662-46675-9_21}.

\bibitem[Mancebo et~al.(2021)Mancebo, Calero, and Garc{\'i}a]{Mancebo2021DoesMaintainabilityRelate}
Javier Mancebo, Coral Calero, and F{\'e}lix Garc{\'i}a.
\newblock Does maintainability relate to the energy consumption of software? {{A}} case study.
\newblock \emph{Softw. Qual. J.}, 29\penalty0 (1):\penalty0 101--127, March 2021.
\newblock ISSN 1573-1367.
\newblock \doi{10.1007/S11219-020-09536-9}.

\bibitem[Maquoi et~al.(2025)Maquoi, Cauz, Vanderose, and Devroey]{Maquoi2025EnergyCodesumptionLeveraging}
J{\'e}r{\^o}me Maquoi, Maxime Cauz, Beno{\^{\i}}t Vanderose, and Xavier Devroey.
\newblock Energy codesumption, leveraging test execution for source code energy consumption analysis.
\newblock In Leonardo Montecchi, Jingyue Li, Denys Poshyvanyk, and Dongmei Zhang, editors, \emph{Proceedings of the 33rd {{ACM}} International Conference on the Foundations of Software Engineering, {{FSE}} Companion 2025, Clarion Hotel Trondheim, Trondheim, Norway, June 23-28, 2025}, pages 1432--1436. ACM, 2025.
\newblock \doi{10.1145/3696630.3728707}.

\bibitem[McCullough et~al.(2011)McCullough, Agarwal, Chandrashekar, Kuppuswamy, Snoeren, and Gupta]{McCullough2011EvaluatingEffectivenessModelbased}
John~C. McCullough, Yuvraj Agarwal, Jaideep Chandrashekar, Sathyanarayan Kuppuswamy, Alex~C. Snoeren, and Rajesh~K. Gupta.
\newblock Evaluating the effectiveness of model-based power characterization.
\newblock In \emph{Proceedings of the 2011 {{USENIX}} Conference on {{USENIX}} Annual Technical Conference}, {{USENIXATC}}'11, page~12, USA, 2011. USENIX Association.

\bibitem[Noureddine(2022)]{Noureddine2022PowerJoularJoularJXMultiPlatform}
Adel Noureddine.
\newblock {{PowerJoular}} and {{JoularJX}}: {{Multi-Platform Software Power Monitoring Tools}}.
\newblock In \emph{18th {{International Conference}} on {{Intelligent Environments}}}, pages 1--4, Biarritz, France, June 2022. IEEE.
\newblock \doi{10.1109/IE54923.2022.9826760}.

\bibitem[O'brien et~al.(2017)O'brien, Pietri, Reddy, Lastovetsky, and Sakellariou]{Obrien2017SurveyPowerEnergy}
Kenneth O'brien, Ilia Pietri, Ravi Reddy, Alexey Lastovetsky, and Rizos Sakellariou.
\newblock A {{Survey}} of {{Power}} and {{Energy Predictive Models}} in {{HPC Systems}} and {{Applications}}.
\newblock \emph{ACM Comput. Surv.}, 50\penalty0 (3):\penalty0 37:1--37:38, June 2017.
\newblock ISSN 0360-0300.
\newblock \doi{10.1145/3078811}.

\bibitem[Ournani et~al.(2021{\natexlab{a}})Ournani, Belgaid, Rouvoy, Rust, and Penhoat]{Ournani2021EvaluatingImpactJava}
Zakaria Ournani, Mohammed~Chakib Belgaid, Romain Rouvoy, Pierre Rust, and Jo{\"e}l Penhoat.
\newblock Evaluating the {{Impact}} of {{Java Virtual Machines}} on {{Energy Consumption}}.
\newblock In {Filippo Lanubile}, {Marcos Kalinowski}, and {Maria Teresa Baldassarre}, editors, \emph{{{ESEM}} '21: {{ACM}} / {{IEEE International Symposium}} on {{Empirical Software Engineering}} and {{Measurement}}, {{Bari}}, {{Italy}}, {{October}} 11-15, 2021}, pages 15:1--15:11, Bari Italy, October 2021{\natexlab{a}}. ACM.
\newblock ISBN 978-1-4503-8665-4.
\newblock \doi{10.1145/3475716.3475774}.

\bibitem[Ournani et~al.(2021{\natexlab{b}})Ournani, Rouvoy, Rust, and Penhoat]{Ournani2021EvaluatingEnergyConsumption}
Zakaria Ournani, Romain Rouvoy, Pierre Rust, and Joel Penhoat.
\newblock Evaluating {{The Energy Consumption}} of {{Java I}}/{{O APIs}}.
\newblock In \emph{{{ICSME}} 2021 - 37th {{International Conference}} on {{Software Maintenance}} and {{Evolution}}}, Proceedings of the 37th {{International Conference}} on {{Software Maintenance}} and {{Evolution}} ({{ICSME}}), pages 1--11, Luxembourg / Virtual, Luxembourg, September 2021{\natexlab{b}}. IEEE.
\newblock \doi{10.1109/ICSME52107.2021.00007}.

\bibitem[Pang et~al.(2016)Pang, Hindle, Adams, and Hassan]{Pang2016WhatProgrammersKnow}
Candy Pang, Abram Hindle, Bram Adams, and Ahmed~E. Hassan.
\newblock What {{Do Programmers Know}} about {{Software Energy Consumption}}?
\newblock \emph{IEEE Soft.}, 33\penalty0 (3):\penalty0 83--89, May 2016.
\newblock ISSN 0740-7459, 1937-4194.
\newblock \doi{10.1109/MS.2015.83}.

\bibitem[Papadakis et~al.(2019)Papadakis, Kintis, Zhang, Jia, Traon, and Harman]{papadakis_mutation_2019}
Mike Papadakis, Marinos Kintis, Jie Zhang, Yue Jia, Yves~Le Traon, and Mark Harman.
\newblock Mutation {Testing} {Advances}: {An} {Analysis} and {Survey}.
\newblock In \emph{Advances in {Computers}}, volume 112, pages 275--378. Elsevier, 2019.
\newblock \doi{10.1016/bs.adcom.2018.03.015}.

\bibitem[Pathania et~al.(2025)Pathania, Bamby, Mehra, Sikand, Sharma, Kaulgud, Podder, and Burden]{Pathania2025CalculatingSoftwaresEnergy}
Priyavanshi Pathania, Nikhil Bamby, Rohit Mehra, Samarth Sikand, Vibhu~Saujanya Sharma, Vikrant Kaulgud, Sanjay Podder, and Adam~P. Burden.
\newblock Calculating {{Software}}'s {{Energy Use}} and {{Carbon Emissions}}: {{A Survey}} of the {{State}} of {{Art}}, {{Challenges}}, and the {{Way Ahead}}.
\newblock In \emph{9th {{IEEE}}/{{ACM International Workshop}} on {{Green}} and {{Sustainable Software}}, {{GREENS}}@{{ICSE}} 2025, {{Ottawa}}, {{ON}}, {{Canada}}, {{April}} 29, 2025}, pages 92--99. IEEE, April 2025.
\newblock \doi{10.1109/GREENS66463.2025.00018}.

\bibitem[Penzenstadler et~al.(2012)Penzenstadler, Bauer, Calero, and Franch]{Penzenstadler2012SustainabilitySoftwareEngineering}
B.~Penzenstadler, V.~Bauer, C.~Calero, and X.~Franch.
\newblock Sustainability in software engineering: A systematic literature review.
\newblock In \emph{16th {{International Conference}} on {{Evaluation}} \& {{Assessment}} in {{Software Engineering}} ({{EASE}} 2012)}, pages 32--41, Ciudad Real, 2012. {Institution of Engineering and Technology}.
\newblock ISBN 978-1-84919-541-6.
\newblock \doi{10.1049/ic.2012.0004}.

\bibitem[Pereira et~al.(2017)Pereira, Couto, Ribeiro, Rua, Cunha, Fernandes, and Saraiva]{Pereira2017EnergyEfficiencyProgramming}
Rui Pereira, Marco Couto, Francisco Ribeiro, Rui Rua, J{\'a}come Cunha, Jo{\~a}o~Paulo Fernandes, and Jo{\~a}o Saraiva.
\newblock Energy efficiency across programming languages: How do energy, time, and memory relate?
\newblock In \emph{Proceedings of the 10th {{ACM SIGPLAN International Conference}} on {{Software Language Engineering}}}, {{SLE}} 2017, pages 256--267, New York, NY, USA, October 2017. Association for Computing Machinery.
\newblock ISBN 978-1-4503-5525-4.
\newblock \doi{10.1145/3136014.3136031}.

\bibitem[Poy et~al.(2025)Poy, Moraga, Garc{\'i}a, and Calero]{Poy2025ImpactEnergyConsumption}
Olivia Poy, M~{\'A}ngeles Moraga, F{\'e}lix Garc{\'i}a, and Coral Calero.
\newblock Impact on energy consumption of design patterns, code smells and refactoring techniques: {{A}} systematic mapping study.
\newblock \emph{Journal of Systems and Software}, 222:\penalty0 112303, April 2025.
\newblock ISSN 0164-1212.
\newblock \doi{10.1016/j.jss.2024.112303}.

\bibitem[Raffin and Trystram(2025)]{Raffin2025DissectingSoftwarebasedMeasurement}
Guillaume Raffin and Denis Trystram.
\newblock Dissecting the software-based measurement of {{CPU}} energy consumption: A comparative analysis.
\newblock \emph{IEEE Transactions on Parallel and Distributed Systems}, 36:\penalty0 96--107, 2025.
\newblock \doi{10.1109/TPDS.2024.3492336}.

\bibitem[Roque et~al.(2025)Roque, Cruz, and Durieux]{Roque2025UnveilingEnergyVampires}
Enrique~Barba Roque, Luis Cruz, and Thomas Durieux.
\newblock Unveiling the {{Energy Vampires}}: {{A Methodology}} for {{Debugging Software Energy Consumption}}.
\newblock In \emph{47th {{IEEE}}/{{ACM International Conference}} on {{Software Engineering}}, {{ICSE}} 2025, {{Ottawa}}, {{ON}}, {{Canada}}, {{April}} 26 - {{May}} 6, 2025}, pages 2406--2418. IEEE, April 2025.
\newblock \doi{10.1109/ICSE55347.2025.00118}.

\bibitem[Schubert et~al.(2012)Schubert, Kostic, Zwaenepoel, and Shin]{Schubert2012ProfilingSoftwareEnergy}
Simon Schubert, Dejan Kostic, Willy Zwaenepoel, and Kang~G. Shin.
\newblock Profiling {{Software}} for {{Energy Consumption}}.
\newblock In \emph{2012 {{IEEE International Conference}} on {{Green Computing}} and {{Communications}}}, pages 515--522. IEEE Computer Society, November 2012.
\newblock \doi{10.1109/GreenCom.2012.86}.

\bibitem[Schuler and Kotsis(2022)]{Schuler2022MANAiIntelliJPlugin}
Andreas Schuler and Gabriele Kotsis.
\newblock {{MANAi}} -- {{An IntelliJ Plugin}} for {{Software Energy Consumption Profiling}}.
\newblock \emph{CoRR}, abs/2205.03120, May 2022.
\newblock \doi{10.48550/ARXIV.2205.03120}.

\bibitem[Shahid et~al.(2017)Shahid, Fahad, Reddy, and Lastovetsky]{Shahid2017AdditivitySelectionCriterion}
Arsalan Shahid, Muhammad Fahad, Ravi Reddy, and Alexey Lastovetsky.
\newblock Additivity: {{A Selection Criterion}} for {{Performance Events}} for {{Reliable Energy Predictive Modeling}}.
\newblock \emph{Supercomputing Frontiers and Innovations}, 4\penalty0 (4):\penalty0 50--65, November 2017.
\newblock ISSN 2313-8734.
\newblock \doi{10.14529/jsfi170404}.

\bibitem[Shapiro and Wilk(1965)]{ShapiroWilk1965}
S.~S. Shapiro and M.~B. Wilk.
\newblock An analysis of variance test for normality (complete samples).
\newblock \emph{Biometrika}, 52\penalty0 (3-4):\penalty0 591--611, 12 1965.
\newblock ISSN 0006-3444.
\newblock \doi{10.1093/biomet/52.3-4.591}.

\bibitem[Simon et~al.(2025)Simon, Ekchajzer, Berthelot, Fourboul, Rince, and Rouvoy]{Simon2025BoaviztAPIBottomUpModel}
Thibault Simon, David Ekchajzer, Adrien Berthelot, Eric Fourboul, Samuel Rince, and Romain Rouvoy.
\newblock {{BoaviztAPI}}: {{A Bottom-Up Model}} to {{Assess}} the {{Environmental Impacts}} of {{Cloud Services}}.
\newblock \emph{SIGENERGY Energy Inform. Rev.}, 4\penalty0 (5):\penalty0 84--90, April 2025.
\newblock \doi{10.1145/3727200.3727213}.

\bibitem[{Spencer Desrochers} et~al.(2016){Spencer Desrochers}, {Chad Paradis}, and {Vincent M. Weaver}]{SpencerDesrochers2016ValidationDRAMRAPL}
{Spencer Desrochers}, {Chad Paradis}, and {Vincent M. Weaver}.
\newblock A {{Validation}} of {{DRAM RAPL Power Measurements}}.
\newblock In {Bruce L. Jacob}, editor, \emph{Proceedings of the {{Second International Symposium}} on {{Memory Systems}}, {{MEMSYS}} 2016, {{Alexandria}}, {{VA}}, {{USA}}, {{October}} 3-6, 2016}, pages 455--470, New York, NY, USA, October 2016. ACM.
\newblock ISBN 978-1-4503-4305-3.
\newblock \doi{10.1145/2989081.2989088}.

\bibitem[Tukey(1949)]{Tukey1949ComparingIndividualMeans}
John~W. Tukey.
\newblock Comparing {{Individual Means}} in the {{Analysis}} of {{Variance}}.
\newblock \emph{Biometrics}, 5\penalty0 (2):\penalty0 99--114, 1949.
\newblock ISSN 0006-341X.
\newblock \doi{10.2307/3001913}.

\bibitem[Welch(1947)]{Welch1947}
B.~L. Welch.
\newblock The generalization of ‘student's’ problem when several different population variances are involved.
\newblock \emph{Biometrika}, 34\penalty0 (1-2):\penalty0 28--35, 01 1947.
\newblock ISSN 0006-3444.
\newblock \doi{10.1093/biomet/34.1-2.28}.

\bibitem[Wilk and Gnanadesikan(1968)]{Wilk1968ProbabilityPlottingMethods}
M.~B. Wilk and R.~Gnanadesikan.
\newblock Probability plotting methods for the analysis for the analysis of data.
\newblock \emph{Biometrika}, 55\penalty0 (1):\penalty0 1--17, March 1968.
\newblock ISSN 0006-3444.
\newblock \doi{10.1093/biomet/55.1.1}.

\bibitem[Xiao et~al.(2025)Xiao, Gao, and Bogner]{Xiao2025EffectivenessMicroservicesTactics}
Xingwen Xiao, Chushu Gao, and Justus Bogner.
\newblock On the {{Effectiveness}} of {{Microservices Tactics}} and {{Patterns}} to {{Reduce Energy Consumption}}: {{An Experimental Study}} on {{Trade-Offs}}.
\newblock In \emph{22nd {{IEEE International Conference}} on {{Software Architecture}}, {{ICSA}} 2025, {{Odense}}, {{Denmark}}, {{March}} 31 - {{April}} 4, 2025}, pages 164--175. IEEE, March 2025.
\newblock \doi{10.1109/ICSA65012.2025.00025}.

\end{thebibliography}

\end{document}